\documentclass[twocolumn,trackchanges]{aastex7}
\usepackage{longtable}

\begin{document}

\title{Correlations Between Kilohertz Quasi-periodic Oscillations and Spectral Parameters from Time-Resolved Spectro-Temporal Analysis of 4U 1728–34}

\author[orcid=0009-0006-0457-2091,sname='Anand']{Kewal Anand}
\affiliation{Department of Physics, Indian Institute of Technology Kanpur, Kanpur Nagar, Uttar Pradesh 208016, India}
\email[show]{kanand@iitk.ac.in}

\author[orcid=0000-0000-0000-0002,sname='Misra']{Ranjeev Misra} 
\affiliation{Inter-University Center for Astronomy \& Astrophysics, Ganeshkhind, Pune 411007, India}
\email{rmisra@iucaa.in}

\author[orcid=0000-0000-0000-0004,sname='Yadav']{J. S. Yadav}
\affiliation{Department of Space, Planetary \& Astronomical Sciences \& Engineering, IIT Kanpur, Kanpur Nagar, Uttar Pradesh 208016, India}
\affiliation{Department of Astronomy \& Astrophysics, Tata Institute of Fundamental Research, Colaba, Mumbai 400005, India}
\email{jsyadav@iitk.ac.in}

\author[orcid=0000-0000-0000-0004,sname='Jain']{Pankaj Jain}
\affiliation{Department of Space, Planetary \& Astronomical Sciences \& Engineering, IIT Kanpur, Kanpur Nagar, Uttar Pradesh 208016, India}
\email{pkjain@iitk.ac.in}

\begin{abstract}
We present a time-resolved analysis of the persistent emission in 4U 1728--34 using AstroSat observations from 2016 to 2019. We detect kilohertz quasi-periodic oscillations (kHz QPOs) during all epochs, with centroid frequencies ranging from $\sim$350 to 1180 Hz, although some detections are of lower significance ($<3\sigma$). We model the simultaneous spectra from the Soft X-ray Telescope and the Large Area X-ray Proportional Counter using a combination of an absorbed disk component (\verb'diskbb'), a blackbody component (\verb'bbodyrad'), a thermal Comptonization model (\verb'thcomp'), and a broad \verb'gaussian' line. From \verb'diskbb' parameters, we estimate the accretion rate and find that all the observations fall into two accretion regimes, namely AR1 and AR2, with accretion rates of $\sim$ $3 \times 10^{16} \, \text{g} \, \text{s}^{-1}$ and $7 \times 10^{16} \, \text{g} \, \text{s}^{-1}$, respectively. Interestingly, we find that for AR1, the lower kHz QPO frequency ($\nu_{\text{L}}$) is always $<$500 Hz, while for AR2 it is $\gtrsim$500 Hz. We found that the spectral index showed no clear correlation with $\nu_{\text{L}}$. For AR1, the coronal electron temperature (kT$_e$) and optical depth ($\tau$) are $\sim$10 and $\sim$5 keV, respectively. In contrast, for AR2, kT$_e$ decreases to $\sim$3 keV and $\tau$ increases to $\sim$12, showing correlations with $\nu_{\text{L}}$, with Spearman's rank correlation coefficients of $–0.78$, $0.71$, respectively. The transition of the spectral parameters at $\nu_{\text{L}} \sim500$ Hz, indicates the existence of a critical QPO frequency that is governed or influenced by the source's accretion state.

\end{abstract}



\section{Introduction} 
X-ray binaries (XRBs) consist of binary star systems of either a neutron star (NS) or a black hole (BH) as a primary and a conventional gaseous star as a secondary. XRBs are mainly classified into three categories based on the mass of their companion star: low-mass X-ray binary (LMXB), intermediate-mass X-ray binary (IMXRB), and high-mass X-ray binary (HMXB). In LMXBs, the mass of the companion star is $\lesssim 1 \,M_\odot$, and the accretion takes place via Roche lobe overflow. HMXBs have donor stars having mass $> 10 \,M_\odot$, and the accretion in such systems is powered by the stellar wind mechanism. On the other hand, the mass of the secondary star in IMXBs lies between $\sim$2--10 M$_\odot$. The NS in LMXBs possesses comparatively a low magnetic field ($< 10^9$ G), and hence, the accreted in-falling matter flow remains largerly unaffated by the magetic field. The accreted matter in NS-HMXB experiences a strong magnetic field ($\sim$10$^{12}$ G) and is funneled onto the poles of the NS through the magnetic field lines \citep[for a review, see][]{lewin1997}.

NS-LMXBs are further classified as Z-type and atoll-type sources based on their tracks on the color-color diagram (CD). Z-type sources are more luminous and, hence, have a higher accretion rate than atoll sources. They radiate at close to the Eddington luminosity (L$_{\text{Edd}}$). In contrast, the luminosity of atoll sources typically remains between $0.01-0.5 \, \text{L}_{\text{Edd}}$. The different spectral states of atoll sources have been labeled as extreme island, island, lower banana, and upper banana on the CD track. On the other hand, the spectral states of Z sources are referred to as horizontal branch (HB), normal branch (NB), and flaring branch (FB) \citep{hasinger1989}.

XRBs show a wide range of spectral and temporal variabilities. They usually make a transition from one spectral state to the other on the time scale of a few hours to months. The temporal variability is well studied in the Fourier domain by computing the power density spectrum/spectra (PDS) of light curves. PDS of XRBs exhibit some broad and narrow features. The broad features ($Q=\frac{\nu_0}{\Delta \nu}<<$ 2) are referred to as broadband noise, whereas narrow features (Q $\gtrsim$ 2) are known as quasi-periodic oscillations (QPOs). QPOs peak at a certain frequency with some definite width and are generally fitted with the Lorentzian function. The centroid frequency at which the power maximizes is referred to as the QPO frequency \citep{klis1989fourier,klis2000millisecond,ingram2019review}. The QPO frequency ranges from a few mHz to kHz in different XRBs. QPOs with centroid frequency $\sim$0.1 to 60 Hz and $\sim$60 to 300 Hz are known as low-frequency QPOs (LF QPOs) and high-frequency QPOs (HF QPOs), respectively. Furthermore, HF QPOs observed in neutron star XRBs have centroid frequencies ranging from around 300 to 1300 Hz and are known as kHz QPOs. The kHz QPOs and/or HF QPOs) are commonly seen as twin QPOs (lower kHz QPO \& upper kHz QPO) in both neutron star and black hole systems (e.g. \cite{van_der_Klis_1997}). 

In BH XRBs, Type-C QPOs are known to show a tight correlation with the spectral parameters, such as the spectral index, electron temperature, mass accretion rate, and luminosity \citep{Sobczak_2000,vignarca2003,Yadav_2016,rawat_2023,dhaka2023}. Other LF QPOs have also shown some connections between their occurrence and spectral state of the source. For instance, Type-B QPOs are typically detected during the soft-intermediate state, whereas Type-A QPOs are rarely observed and, when observed, usually appear in the soft state shortly after the transition from the hard to the soft state \citep{Belloni2016}. 

A similar trend, exhibiting a one-to-one correlation between the occurrence of LF QPOs and the spectral state, has also been observed in NS systems. In particular, in Z sources, LF QPOs are typically observed in HB, NB, and FB, which are respectively associated with horizontal branch oscillations, normal branch oscillations, and flaring branch oscillations. These LF QPOs follow a well-defined sequence along the Z track, analogous to Type-C, Type-B, and Type-A QPOs observed in the hardness-intensity diagram \citep{Casella_2005,Ingram2019}.

Another important timing feature in NS XRBs is the kHz QPO. The first kHz QPO was discovered in a Z source SCO-X1 from the observation made by the \textit{Rossi X-ray Timing Explorer (RXTE)} in 1996 \citep{van1996discovery,Berger_1996}. The kHz QPOs are often observed at lowest inferred accretion rate \citep{vanderklis2000}. In atoll sources, kHz QPOs are detected when the source lies in lower banana branch which is closest to the island state \citep{Ford_1998}. In Z sources, kHz QPOs are observed in horizontal branch and when source makes transition from horizontal to the normal branch. Most timing and spectral properties of these sources depend upon their position on Z or atoll track \citep{Homan_2002}. The kHz QPOs, especially the upper kHz QPOs, observed in these sources are positively correlated with the CD parameter (S$_s$ or S$_z$) \citep{Mendez1999,vanStraaten2000,belloni2005}. \cite{Ribeiro2017} showed that the kHz QPOs in 4U 1636-536 are correlated with the spectral parameters of the source. They found that the electron temperature and the spectral index are anti-correlated, whereas the optical depth is positively correlated with the frequency of kHz QPOs. Besides these kHz QPOs, LF QPOs are altogether observed in these NS sources. Several reports on the correlation between kHz, LF QPOs, and broad-band noise exist in the literature \citep{strohmayer1996millisecond,Miller_1998,Di_Salvo_2001}, which hints towards common origin of these quasi-periodicities from some yet unknown or poorly understood underlying phenomena.

4U 1728--34 is an atoll source and one of the most studied bursting NS-LMXBs. It exhibits frequent thermonuclear bursts (Type I X-ray bursts) with a duration of $\sim$20 s, and is therefore also known as an X-ray burster \citep[e.g.,][]{Galloway_2008,zhang2016,Chauhan_2017}. The distance to the source is $\sim$5 kpc, as estimated from studies of thermonuclear bursts \citep{Di_Salvo_2000}. The hydrogen column density along the line of sight to this source is estimated to be $\sim$2--4 $\times 10^{22}$ cm$^{-2}$ \citep{hoffman1979,Sleator_2016,Bostanci_2023}.

 In previous work, \cite{Anand_2024} analyzed one of the AstroSat/LAXPC observations (Obs. ID: 9000000362) of 4U 1728-34 and reported multiple sets of QPO triplets, consisting of a LF QPO and twin kHz QPOs. The observed QPOs showed a remarkable correlation with each other, which is consistent with the relativistic precesssion model \citep[RPM;][]{stella_1999}. Considering LF QPO frequency as twice the nodal precession frequency in RPM, they got constraints on the mass and moment inertia of the NS to be $\sim1.9 \, \text{M}_\odot$ and $\sim2 \times 10^{45} \,\text{g}\,\text{cm}^2$, respectively. 
 
 In this work, we have used all available data from LAXPC and SXT instruments onboard AstroSat for studying the spectro-timing properties of the persistent emission in 4U 1728--34. We have perfomed a time-resolved spectral analysis of persistent emission and studied correlations between the kHz QPOs and spectral parameters.    

\section{OBSERVATIONs, INSTRUMENTS \& DATA REDUCTION}
AstroSat carries four science payloads onboard, namely, Ultra Violet Imaging Telescope \citep[UVIT;][]{Tandon_2017},  Soft X-ray Telescope \citep[SXT;][]{singh_2017}, Large Area X-ray Proportional Counter \citep[LAXPC;][]{yadav2016large,Antia_2017}, Cadmium Zinc Telluride Imager \citep[CZTI;][]{rao2017}. AstroSat observed 4U 1728--34 on several occasions from 2016 to 2019. The details of the AstroSat observations used in this work have been tabulated in  Table~\ref{tab:table_1}. The data are public and can be downloaded from the AstroSat Science Data Archive\footnote{\url{https://astrobrowse.issdc.gov.in/astro_archive/archive/Home.jsp}}.
\begin{table}
	\centering
	\caption{The detail of the AstroSat observations of 4U 1728-34 used in this work.}
	\label{tab:table_1}
	\begin{tabular}{|c|c|c|c|c|}
		\hline
		Obs-\# & Obs. ID & Obs. Start Date & \multicolumn{2}{c|}{Exp. Time (ks)}   \\
	     \cline{3-5}
		& & (YYYY-MM-DD) & LAXPC & SXT  \\
		\hline
		Obs-1 & 9000000362 & 2016-03-07 & 56.1 & 18.0  \\
		Obs-2 & 9000000578 & 2016-08-05 & 19.3 & 9.79  \\
		Obs-3 & 9000001904 & 2018-02-19 & 7.64 & 2.93  \\
		Obs-4 & 9000002234 & 2018-07-17 & 10.2 & 7.59   \\
		Obs-5 & 9000002254 & 2018-07-25 & 38.6 & 0   \\
        Obs-6 & 9000002268 & 2018-08-01 & 7.45 & 2.00  \\
        Obs-7 & 9000002890 & 2019-05-04 & 84.1 & 0  \\
	    Obs-8 & 9000003134 & 2019-08-29 &81.1  & 35.3  \\
		\hline
		
	\end{tabular}
\end{table}
\subsection{LAXPC}   
LAXPC consists of three nominally identical but independent detectors, namely LAXPC10, LAXPC20, and LAXPC30, with a total large effective area of $\sim6000 \,\text{cm}^2$. Each LAXPC unit has 5 anode layers, each with 12 anode cells. LAXPC uses a large detection volume filled with Xenon gas with a small percentage of methane gas ($\sim$10 \%) at about 2 atm pressure. In the LAXPC instrument, the data is collected in two different modes, namely broad band counting (BBC) and event analysis (EA) modes. In the EA mode, each LAXPC unit works in the energy range of 3--80 keV with a moderate spectral resolution of $\sim$12 \% at 30 keV. LAXPC records the arrival of each X-ray photon with a time resolution of $10\, \mu s$ and a dead-time of $\sim$42 $\mu s$ \citep{yadav2017large,Antia_2017}. In this work, we used the EA mode data because this mode contains all the information required for timing and spectral analyses.     

We used the most recent \verb'LAXPCsoftware'\footnote{\url{http://astrosat-ssc.iucaa.in/laxpcData}}, an LAXPC data reduction software provided by the instrument team, to get useful science products. \verb'LAXPCsoftware' converts Level-1 data into Level-2 and is also used to extract the light curves, PDS, and energy spectra using specific commands and user-intended flags. The detailed instrument calibration and the background modeling have been discussed by \cite{Antia_2017} and \cite{Antia2021}.

\subsection{SXT}
SXT is a focusing instrument that exploits the principle of grazing incidence to focus the soft X-rays. The SXT optics uses a set of coaxial and confocal shells of gold-coated conical foil mirrors arranged in nearly Wolter-I geometry with a geometric area of $\sim$ 250 \,cm$^{2}$. The cooled Charge Coupled Device (CCD) is placed at the focal plane (focal length 2 m). SXT works in the energy range of 0.3-8.0 keV with an energy resolution of 90 eV and an effective area of $\sim$ 90 cm$^2$ at 1.5 keV. There are four different modes of SXT operation, namely, photon counting (PC), fast window (FW), bias map (BM), and calibration (CB) modes. The science data are contained in PC and FW modes only. We have used PC mode data for the analysis in this work since the data are collected from all of the pixels of CCD in this mode. The time resolution of the PC mode is 2.3775 s \citep{singh_2017,singh2022}. 

The Level 1 data are processed by the SXT payload operation center to create the Level 2 data. A user can download the Level 1 data directly from the ASDA\footnote{\url{http://astrosat-ssc.iucaa.in/laxpcData}} website and then convert them into Level 2 data by the SXT pipeline software \footnote{\url{piplinehttp://astrosat-ssc.iucaa.in/sxtData}}. The cleaned-event file of Level 2 data is read in \verb'Xselect', which is a preinstalled tool of \verb'HeaSoft',  and light curves, energy spectra, and images are extracted. The maximum persistent emission count rate was $\sim$ 30 counts/s; therefore, the pile-up correction was not needed, and hence, we selected a circular source region of radius ranging from 6.5 to 14 arcmin for all epochs. The latest  response matrix file (RMF), ancillary response file (ARF), and background files provided have been used for the analysis. The ARF needs to be corrected based on the source region selected. Therefore, we used the ARF generation tool, \verb'sxtARFModule'\footnote{\url{https://www.tifr.res.in/~astrosat_sxt/dataanalysis.html}}, to generate the corrected ARF file. For grouping the SXT spectrum, ARF, RMF, and backgrouund files, we used \verb'ftgrouppha' of FTOOLS with an optimal binning scheme.

\section{Data Analysis}
\subsection{Temporal Analysis}
\begin{figure}
	
		\includegraphics[width=\columnwidth]{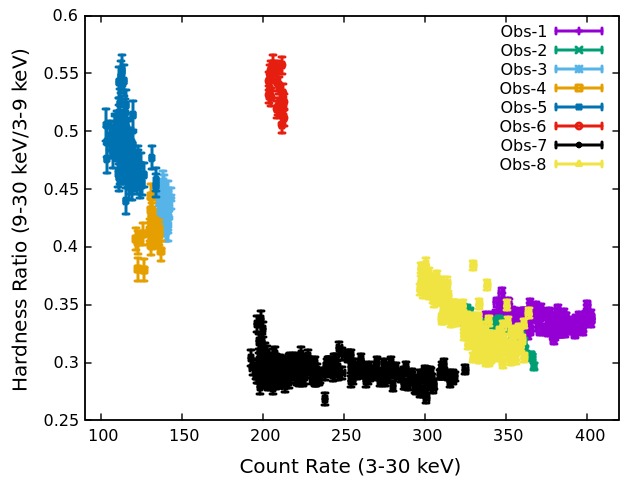}
		\caption{Figure shows the HID of 4U 1728--34 using LAXPC20 data from all eight AstroSat observations described in this work.}
		\label{fig:hid}
	
\end{figure}

For the timing analysis, we used the LAXPC data. There are nine AstroSat observations of 4U 1728--34 in total from 2016 to 2019. One of the observations was excluded from our study due to unusual features in the light curve, which may be attributed to some instrumental effects. Figure~\ref{fig:hid} shows the HID plotted using data from LAXPC20 for all eight remaining epochs. During these epochs, the source underwent the spectral transitions, tracing four different loci on the HID. The average count rate for Obs-1, Obs-2, and Obs-8 is $\sim$ 350 counts s$^{-1}$, while for Obs-7 the count rate is around 250 counts s$^{-1}$. The hardness ratio for the former set of observations is about 0.35, whereas for Obs-7 it is slightly lower ($\sim$ 0.3). For Obs-3, Obs-4, and Obs-5, which correspond to comparatively harder states, the count rate is $\sim$120 counts s$^{-1}$ with a hardness ratio of $\sim$0.45. On the other hand, Obs-6 appears to lie in an intermediate state, with an average count rate and hardness ratio of $\sim$200 counts s$^{-1}$ and $\sim$0.52, respectively. It is clear from HID that the source is in a softer state during observations 1, 2, 7, and 8 (likely in the banana branch), as compared to observations 3, 4, 5, and 6, during which the source appears to be in the island state, closer to the banana branch.

\begin{figure*}
	\centering
	\begin{minipage}{0.30\textwidth}
		\includegraphics[angle=270,width=\textwidth]{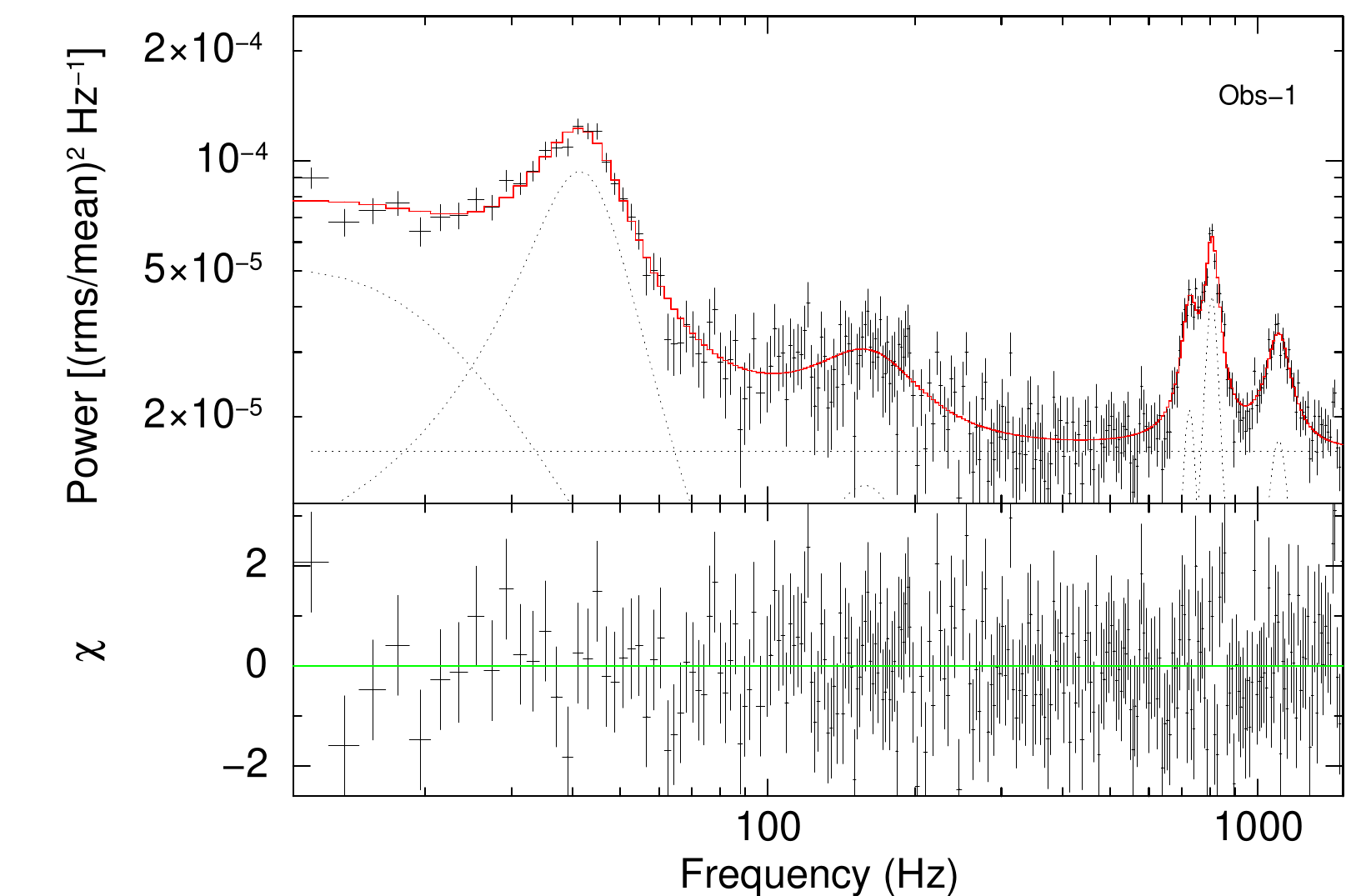}
		\centering {\small (a)}
	\end{minipage}
	\begin{minipage}{0.30\textwidth}
		\includegraphics[angle=270,width=\textwidth]{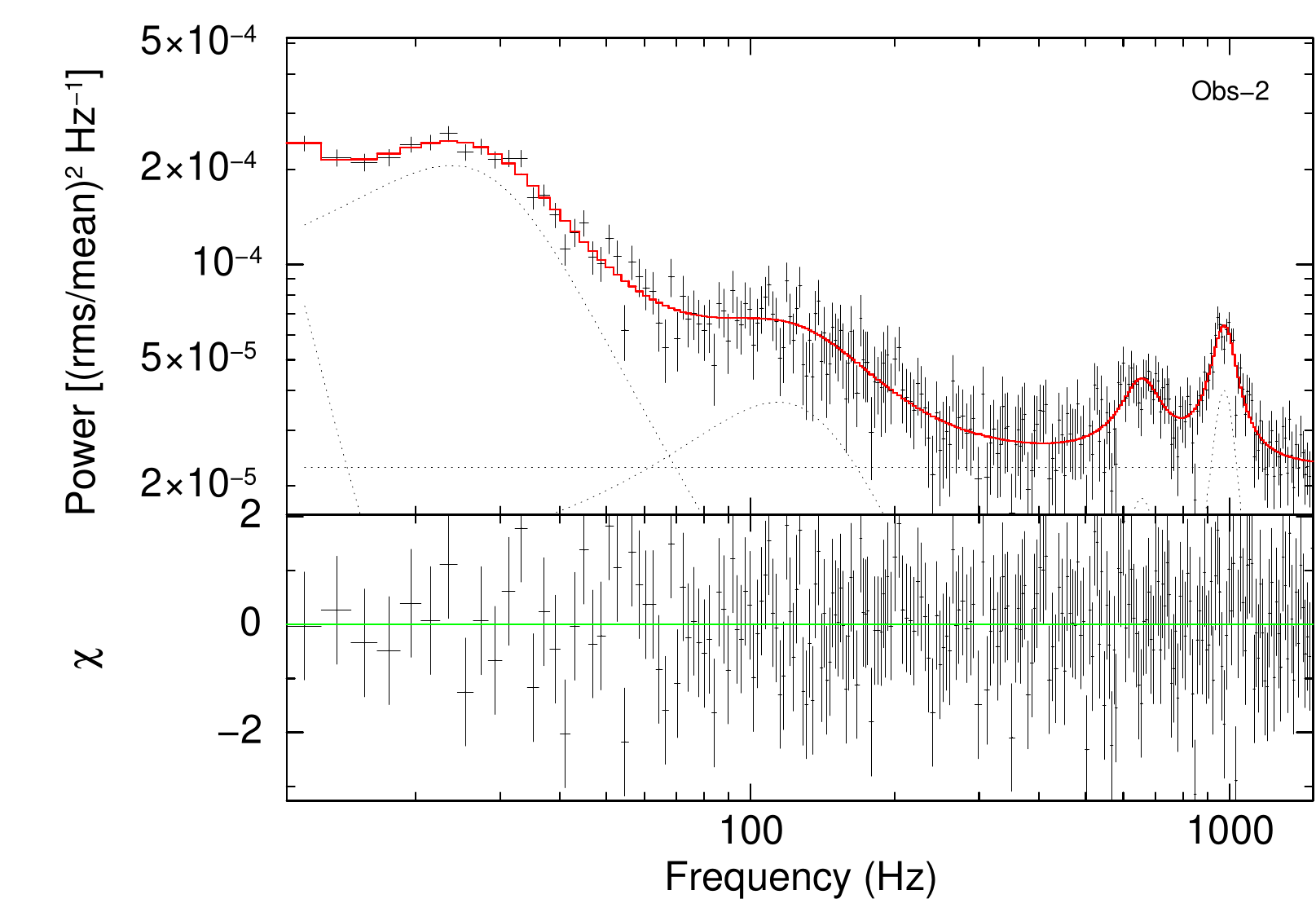}
		\centering {\small (b)}
	\end{minipage}
	\begin{minipage}{0.30\textwidth}
		\includegraphics[angle=270,width=\textwidth]{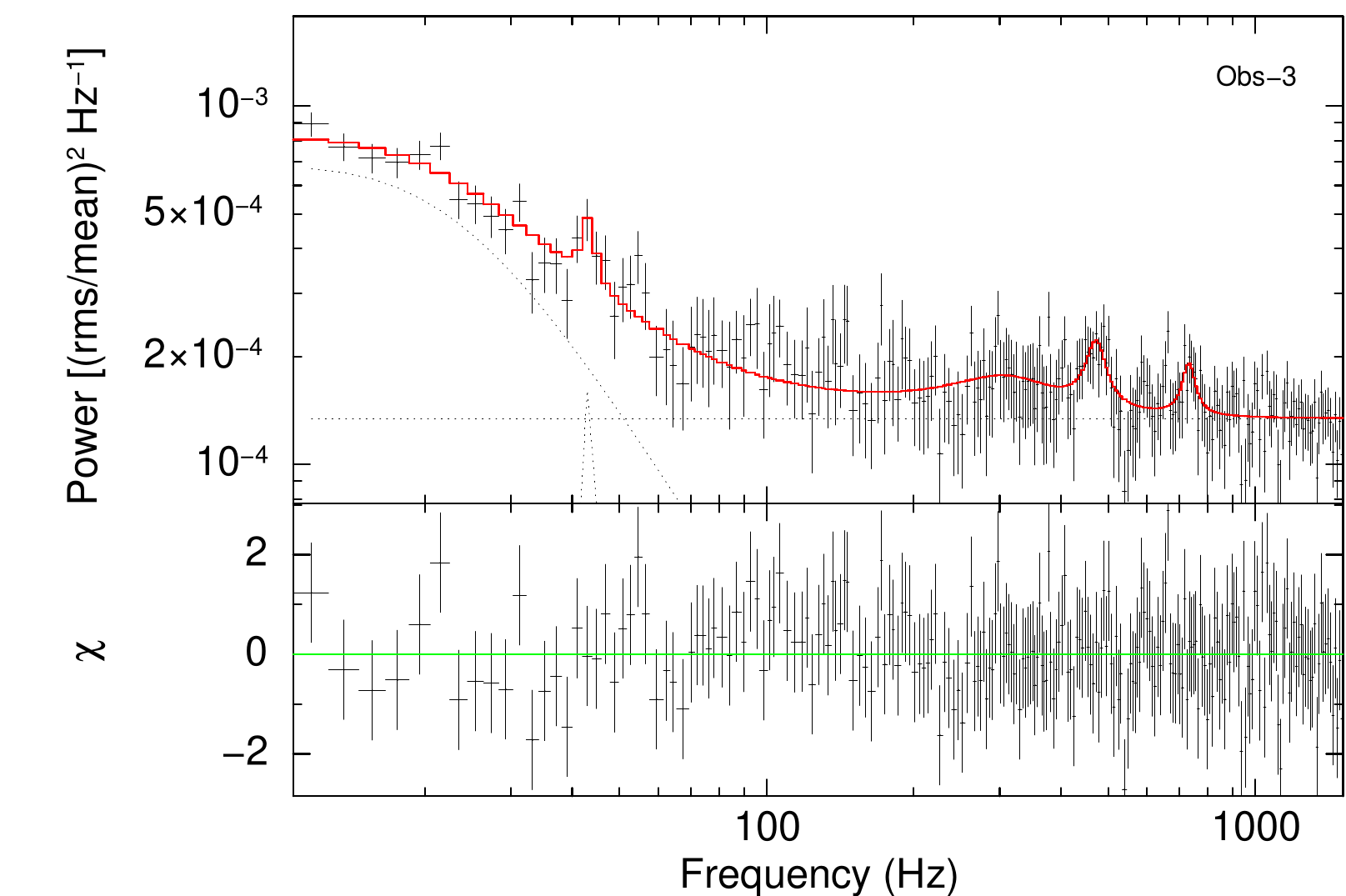}
		\centering {\small (c)}
	\end{minipage}
	
	\vspace{5pt}
	
	\begin{minipage}{0.30\textwidth}
		\includegraphics[angle=270,width=\textwidth]{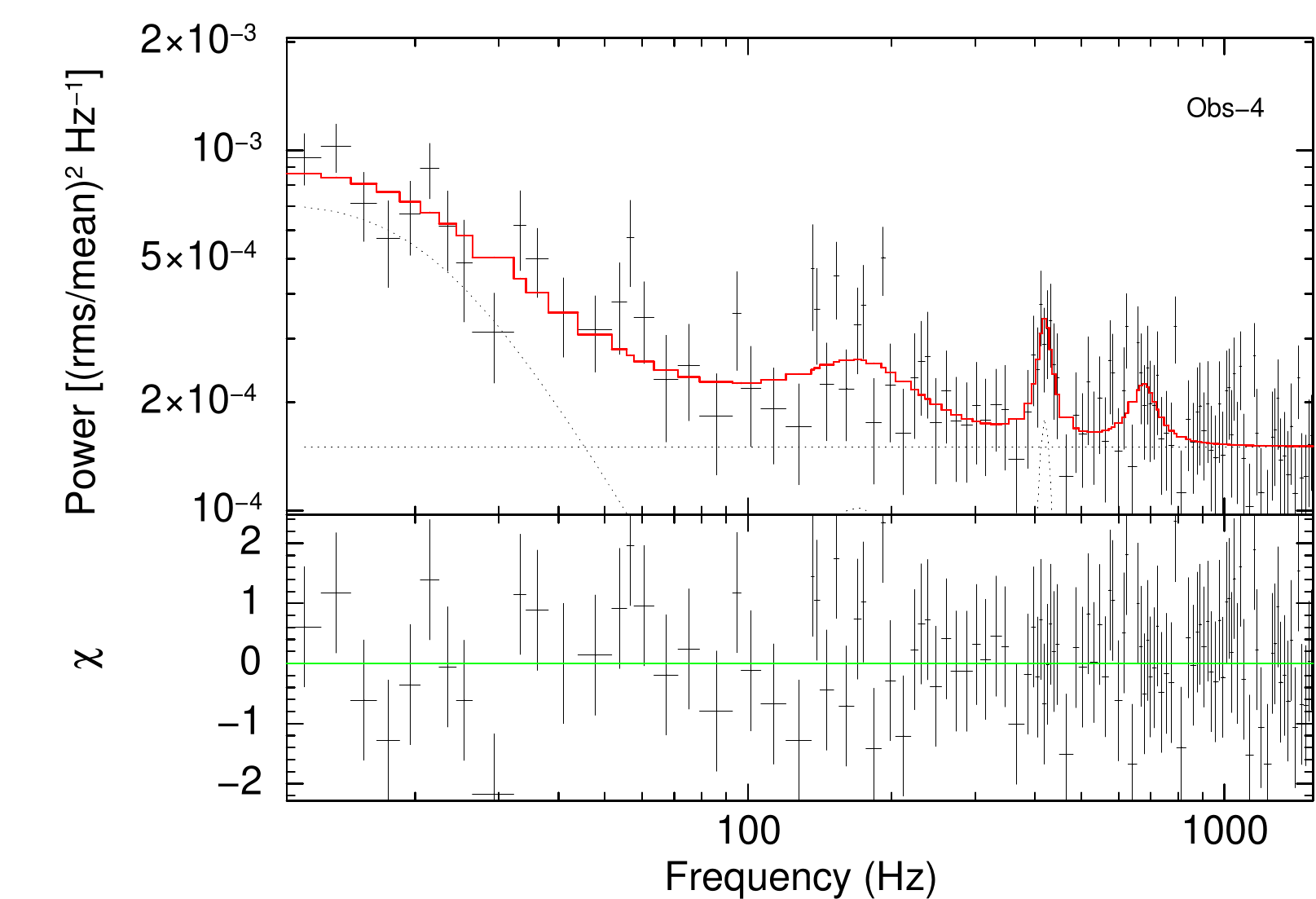}
		\centering {\small (d)}
	\end{minipage}
	\begin{minipage}{0.30\textwidth}
		\includegraphics[angle=270,width=\textwidth]{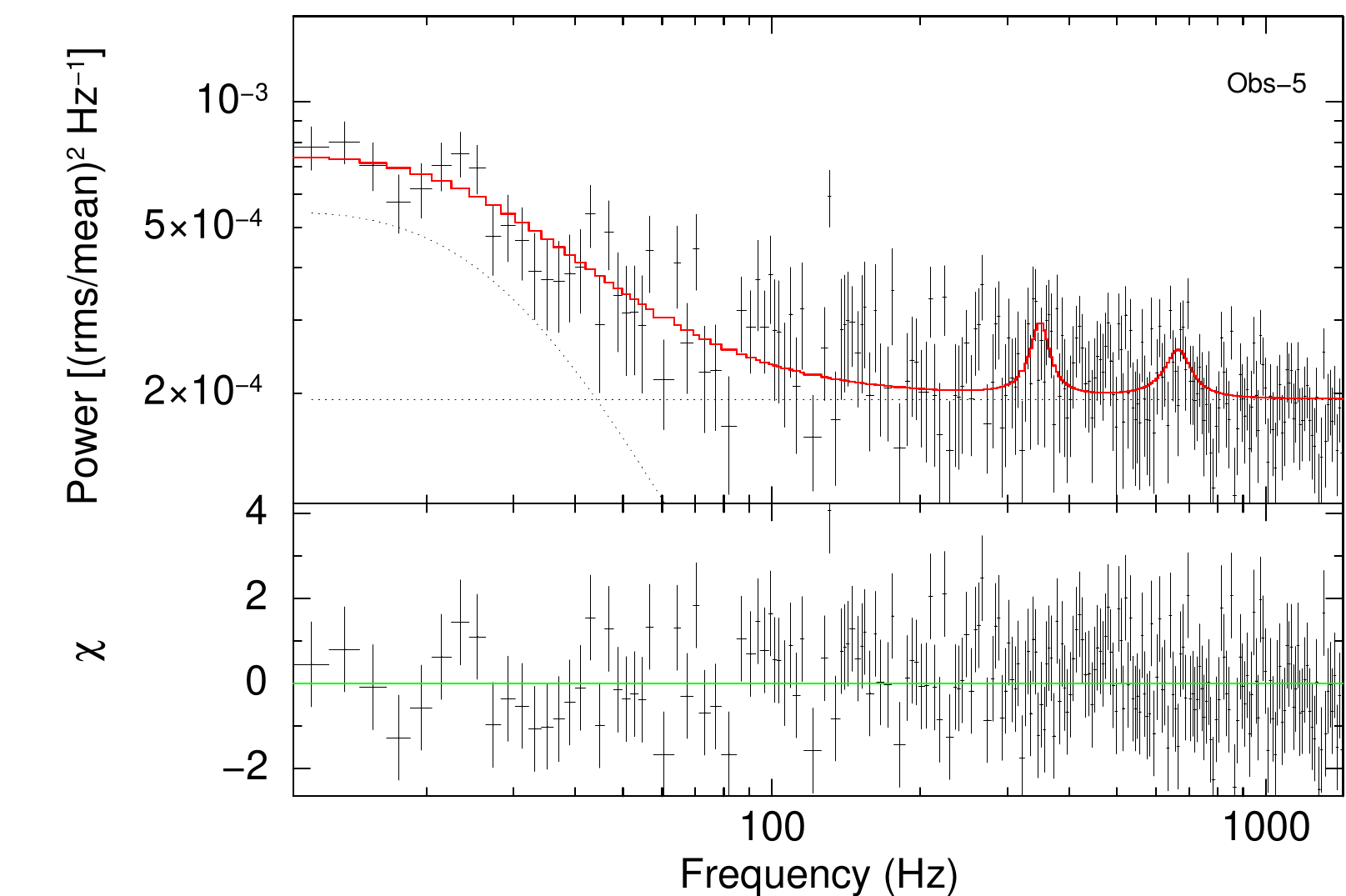}
		\centering {\small (e)}
	\end{minipage}
	\begin{minipage}{0.30\textwidth}
		\includegraphics[angle=270,width=\textwidth]{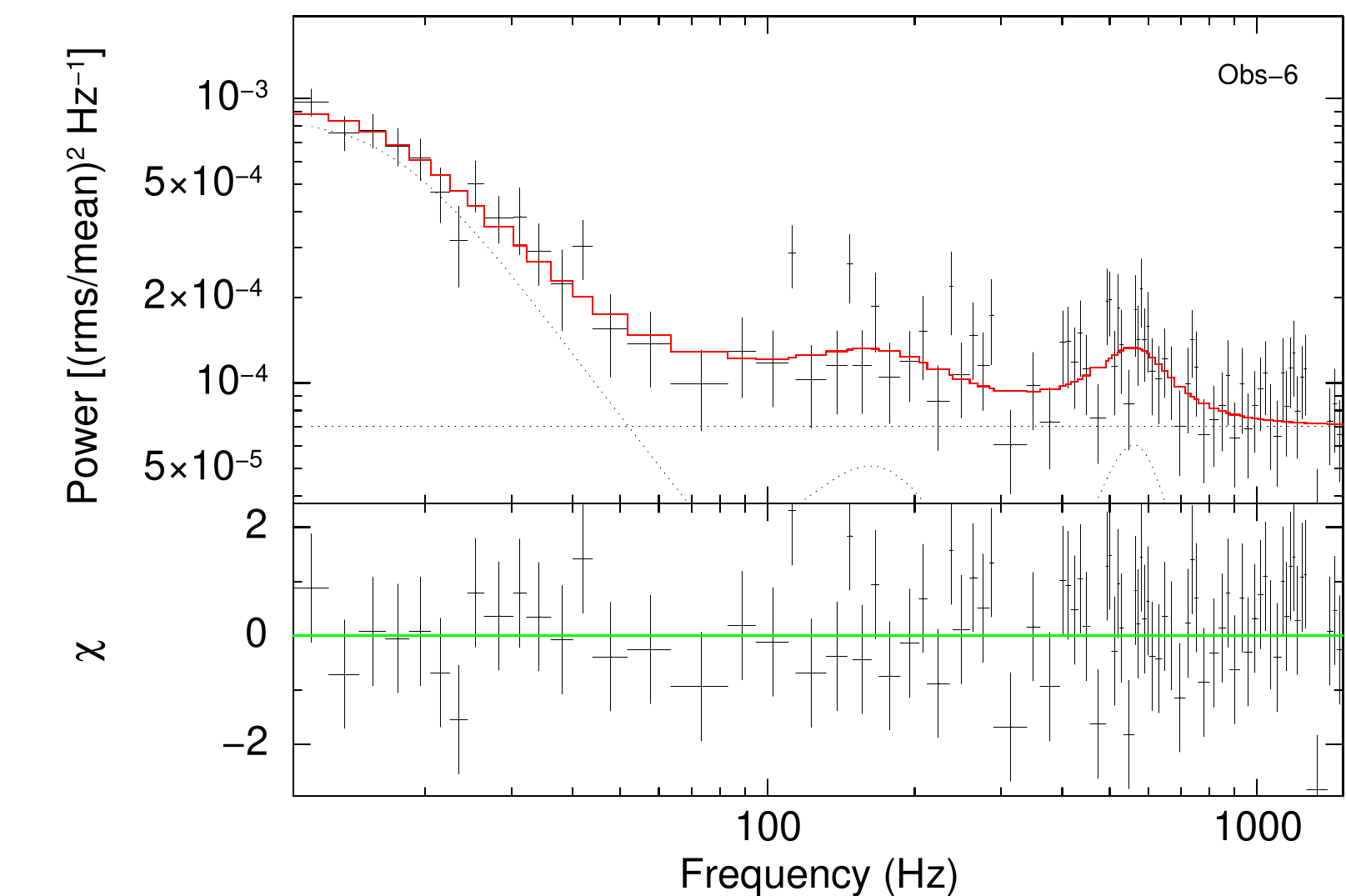}
		\centering {\small (f)}
	\end{minipage}
	
	\vspace{5pt}
	
	\begin{minipage}{0.30\textwidth}
		\includegraphics[angle=270,width=\textwidth]{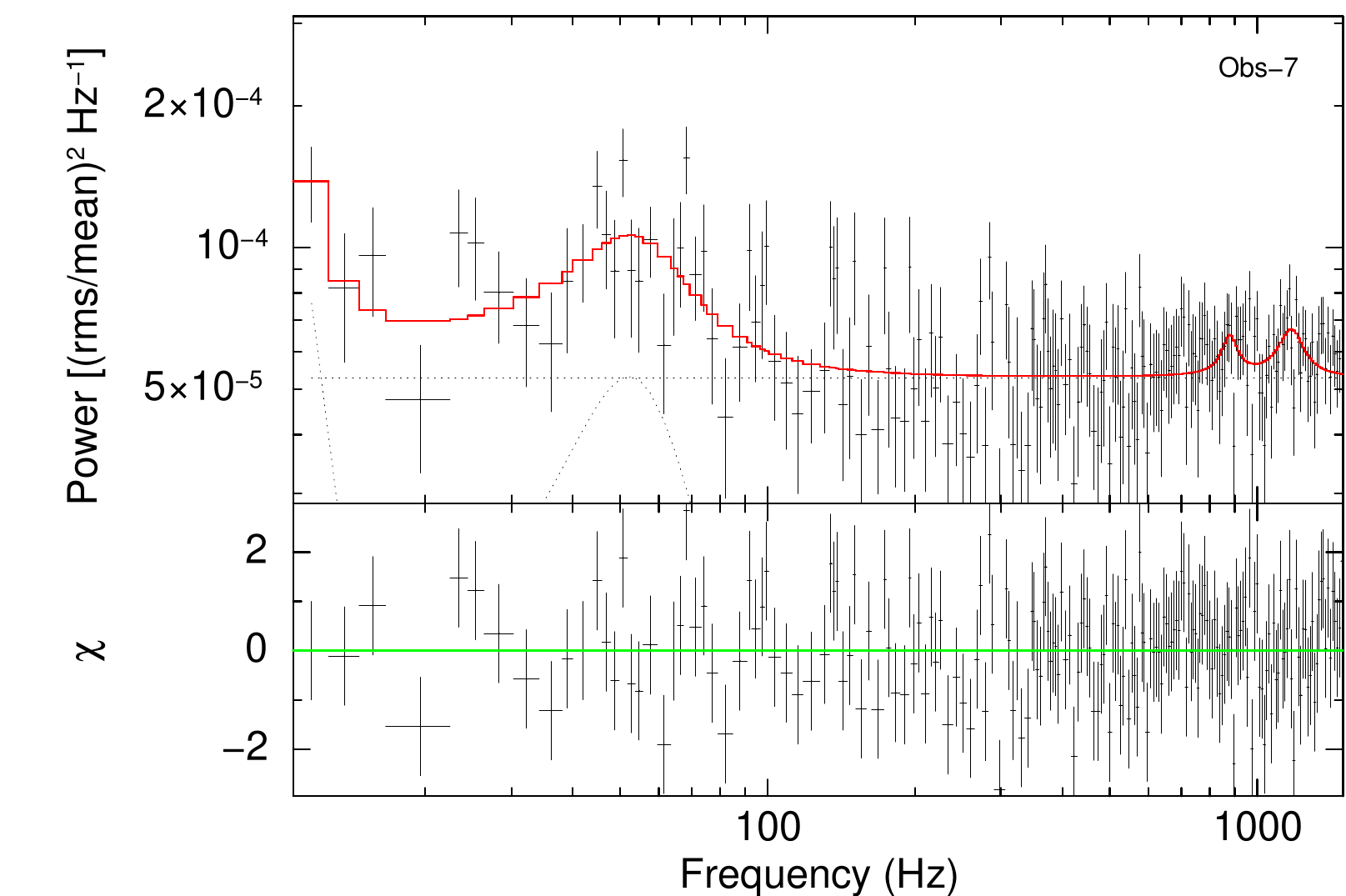}
		\centering {\small (g)}
	\end{minipage}
	\begin{minipage}{0.30\textwidth}
		\includegraphics[angle=270,width=\textwidth]{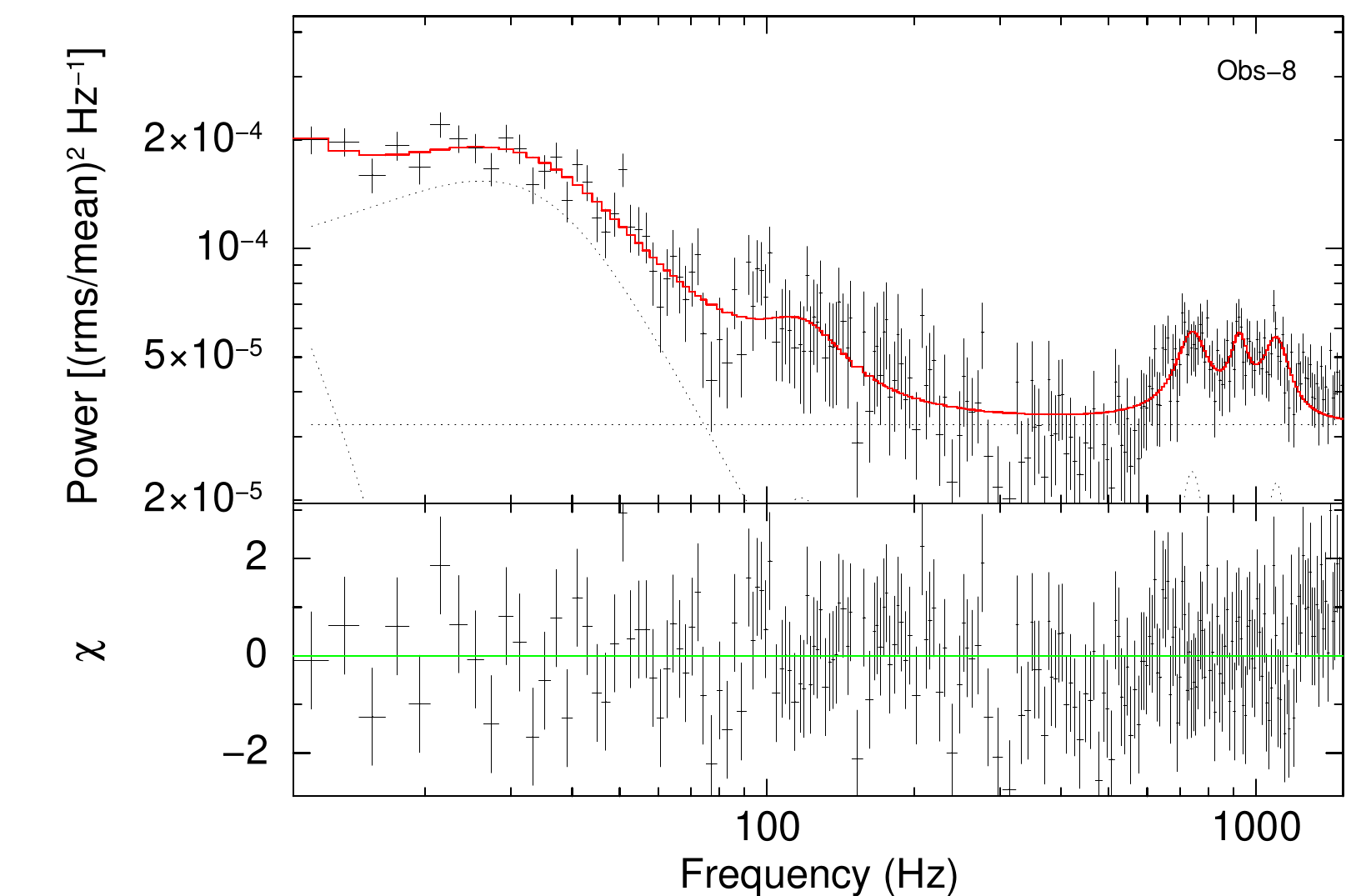}
		\centering {\small (h)}
	\end{minipage}

	\caption{The figure shows the 3–30 keV power density spectra of the persistent emission in the 10–1500 Hz frequency range for all individual observations. Panels (a)–(h) correspond to Obs-1 through Obs-8, respectively.}
	\label{fig:pds_all}
\end{figure*}

The goal of this work is to investigate the properties of kHz QPOs during persistent emission and to look for their correlations with the spectral parameters and the spectral states of the source. Therefore, we removed thermonuclear bursts from the data and Fourier transformed the persistent emission light curves to generate PDS using the command \verb'laxpc_find_freqlag'. Figure~\ref{fig:pds_all} shows the time-averaged PDS for each individual observation. At least one feature can be seen at the kHz region in every PDS. The PDS were fitted with 3--6 Lorentzians, one Lorentzian fixed at 10 Hz to account for a feature at lowest frequencies, and a power-law with zero index, which takes care of a Poisson noise residual at higher frequencies. To generate PDS, we used all three LAXPC units for the first three observations (i.e., Obs-1, 2, and 3) since LAXPC10 and LAXPC30 had some issues\footnote{\url{https://www.tifr.res.in/~astrosat_laxpc/astrosat_laxpc.html}} after March 2018, we have used only LAXPC20 data for rest of the other observations.
  
To see the time evolution of the QPO frequencies and to perform the time-resolved spectroscopy, following \cite{Anand_2024}, we divided some of the observations into multiple time segments. For each segment, we computed the PDS and fitted it using Lorentzians and a zero-indexed power-law. The effective LAXPC exposure time, centroid frequency ($\nu$), width (FWHM; $\Delta \nu$), significance, and fractional root-mean-square (RMS) of the kHz QPOs or high-frequency Lorentzians are summarized in Table~\ref{tab:table_2} (see Appendix~\ref{app:appA}), along with the best-fit statistics. The significance ($s$) of the QPOs is defined as the ratio of the QPO normalization and the lower bound on its 1$\sigma$ error. Significant kHz QPO detections ($s \gtrsim 3$) were found in most segments, although some segments yielded a lower significance. In cases where the QPO width could not be constrained, the upper limit of the width was tied to values corresponding to  Q-factors of 10 and 7 for the lower and upper kHz QPOs, respectively. These values are approximately similar to the Q-factor values of the lower and upper kHz QPOs in the time-averaged PDS of Obs-1.

\subsection{Spectral Analysis}	
\label{sec:sec3.2}
We performed time-resolved spectroscopy of the persistent emission using data from both the SXT and LAXPC instruments on board AstroSat. For each time segment defined for the LAXPC data, we extracted the corresponding SXT light curves and spectra using the same good-time intervals. The SXT and LAXPC20 spectra were then modeled simultaneously in the X-ray spectral fitting package \texttt{XSPEC} (version 12.12.0). For each observation and segment, except for a few cases where there was no SXT exposure time (e.g., Seg-14 of Obs-1, Obs-5, Obs-7, and Seg-3 of Obs-8), we used SXT and LAXPC20 spectra in the energy ranges of 0.7–7 keV and 5–20 keV, respectively. It is recommended to apply gain correction to the SXT spectra; therefore, we used the \verb'gain fit' command with the slope fixed at unity. The gain offset was found to be approximately between $\sim$--1.7 and 80 eV. To account for uncertainties in the LAXPC background estimation and response calibration, we included a 3 \% systematic error in the spectral modeling, as recommended by the instrument team. For the segments or observations where there was no SXT exposure time, we fitted only the LAXPC20 spectra in the energy range of 5–20 keV. Since the hydrogen column density (nH) is sensitive to lower energies ($\lesssim$ 1 keV), it cannot be constrained with the LAXPC data alone. Therefore, we fixed nH to an average value of 2.4 $\times \, 10^{22}$ atoms cm$^{-2}$ during spectral fitting wherever it was not constrained.
\begin{figure*}
	\centering
	\begin{minipage}{0.45\textwidth}
		\includegraphics[angle=270,width=\textwidth]{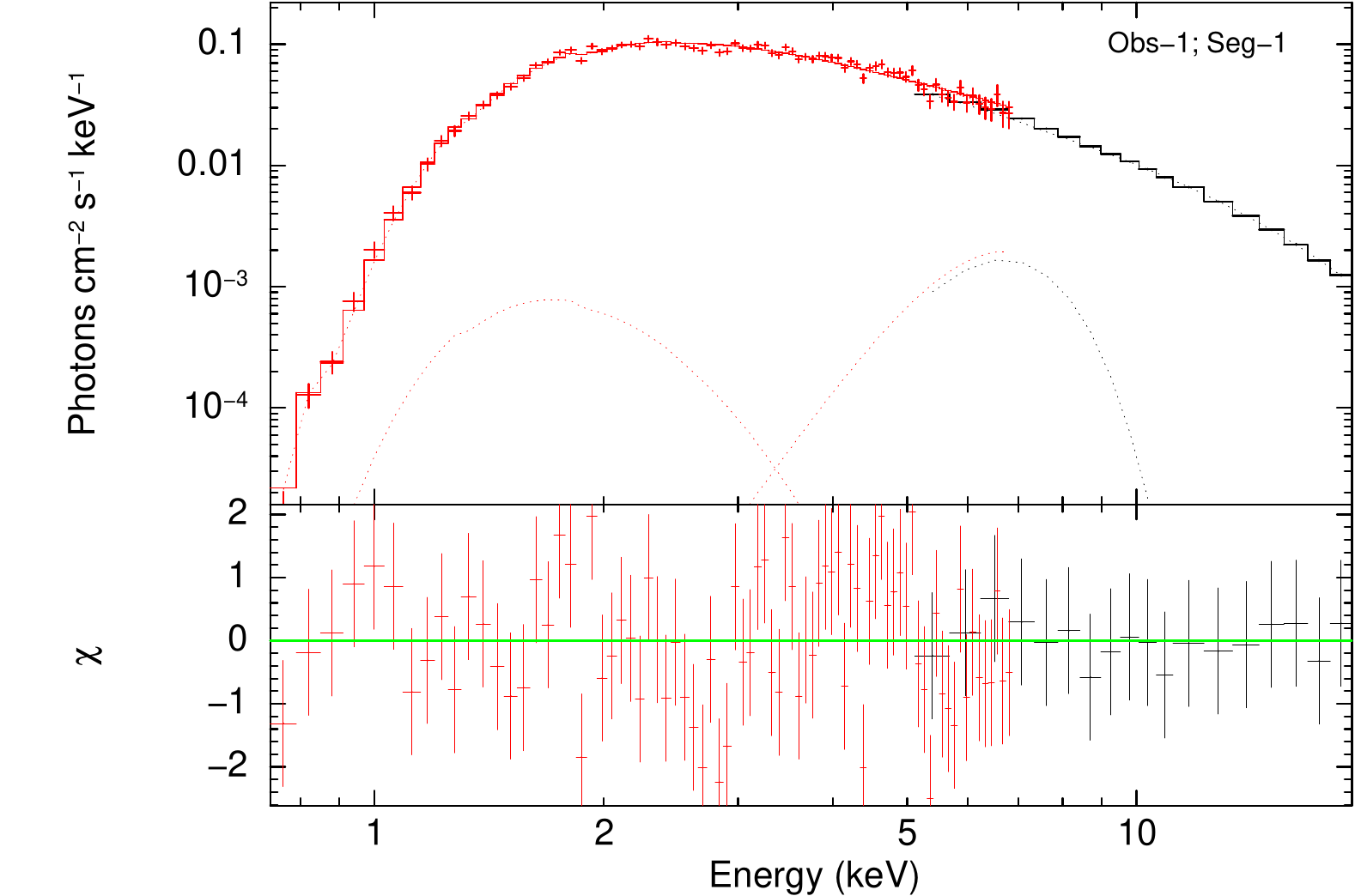}
		\centering {\small (a)}
	\end{minipage}
	\hspace{0.05\textwidth} 
	\begin{minipage}{0.45\textwidth}
		\includegraphics[angle=270,width=\textwidth]{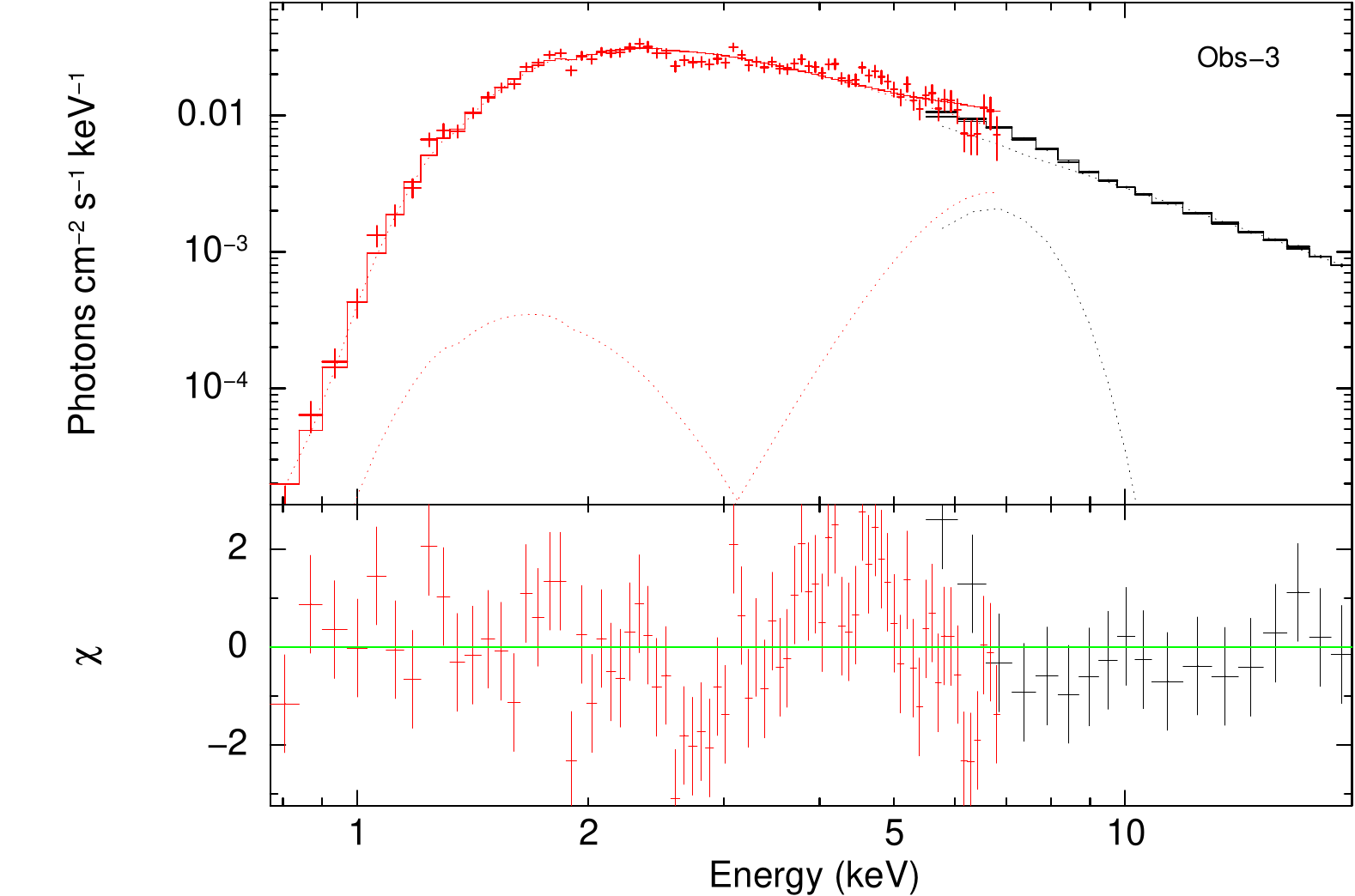}
		\centering {\small (b)}
	\end{minipage}
\caption{The 0.7–20.0 keV SXT+LAXPC20 spectra of Seg-1 of Obs-1 (left) and Obs-3 (right). The spectra have been fitted with the model constant*tbabs*(thcomp*diskbb + bbodyrad + gaussian). The red data points represent the SXT spectra, while the black points are the LAXPC ones.}
\label{fig:spectra}

\end{figure*}

The combined SXT+LAXPC20 spectra for individual segments were fitted using the model \verb'tbabs*(thcomp*diskbb + bbodyrad + gaussian)' in \verb|XSPEC|. A broad \verb|Gaussian| component was required to achieve a good fit for Obs-3, Obs-4, and Obs-8. Although its contribution was negligible in Obs-1, we retained this component in that observation as well to maintain consistency in the spectral modeling across all datasets. During the spectral fitting, the Gaussian line energy was allowed to vary from 6.6--6.9 keV to account for the ionized Fe emission line. In the literature, the Fe fluorescence emission feature in this source has been well constrained and observed to be centered within this range \citep{Di_Salvo_2000,Piraino_2000,Kashyap_2025}. In most of the segments, particularly in Obs-8, the upper limit on the line energy was found to lie between 6.62 and 6.69 keV. Therefore, we fixed the line energy to 6.7 keV for all observations. Additionally, the width of the Gaussian component was also fixed at 0.9 keV wherever it was not constrained.

The following describes the general procedure adopted for modeling the spectrum of each segment. The \verb|Constant| component was used to account for the relative normalization between the two instruments during simultaneous fitting. It was fixed to unity for the LAXPC data and allowed to vary for the SXT data. Interstellar absorption was modeled using \verb|Tbabs| \citep{Wilms_2000}. The Comptonized continuum from the corona was modeled with \verb|Thcomp| \citep{andrzej2020}, which is characterized by four parameters: the spectral index ($\Gamma$), electron temperature (kT$_e$), covering fraction ($f_{\text{cov}}$), and redshift ($z$). The covering fraction was fixed at unity to represent complete Comptonization of the seed photons. The \verb|Thcomp| model allows for the calculation of the Thomson optical depth ($\tau$) by setting the spectral index parameter to a negative value, in which case the optical depth is given by its absolute value \citep{andrzej2020}. Since \verb|Thcomp| is a convolution model, the energy-binning command \textit{``energies 0.01 1000.0 1000 log''} was used to extend the energy range appropriately, as recommended by the model developers.

The accretion disk around the neutron star was modeled using the \verb|Diskbb| model \citep{mitsuda1984}, which describes a geometrically thin, optically thick, multi-temperature blackbody disk. This model has two parameters: the inner disk temperature (T$_{\text{in}}$) and the normalization (N$_{\text{disk}}$), which is defined as:

\begin{equation}
\text{N}_{\text{disk}} = \left( \frac{R_{\text{in}}^{*}}{D_{10}} \right)^2 \cos \theta,
\end{equation}

Here, R$_{\text{in}}^{*}$ is the observed inner disk radius in km, D$_{10}$ is the distance to the source in units of 10 kpc, and $\theta$ is the inclination angle. The actual inner disk radius ($R_{\text{in}}$) is obtained from the observed radius by taking into account for the color correction factor ($\kappa$), which compensates for deviations from a perfect blackbody spectrum due to radiative transfer effects in the disk atmosphere. Since $\text{N}_{\text{disk}}$ could not be constrained during spectral fitting, it was fixed to a value corresponding to an inner radius R$_{\text{in}}^{*}$ assumed to be proportional to the QPO radius (R$_{\text{QPO}}$), which was estimated from the upper kHz QPO frequency ($\nu_{\text{U}}$) as follows:

\begin{equation}
	R_{\text{QPO}}=(GM)^{1/3}\left[\frac{1}{2\pi \nu_{\text{U}}}-\left(\frac{I}{M}\right)\frac{2\pi \nu_{s}}{c^2}\right]^{2/3},
\end{equation}
where M, I/M, and $\nu_s$ are mass, the ratio of moment of inertia to the mass, and the spin frequency of the NS, respectively. These values are taken from \cite{Anand_2024}. The remaining symbols have their usual meanings. Thus, the observed inner disk radius was then calculated using:

\begin{equation}
R_{\text{in}}^{*} = \frac{R_{\text{in}}}{\xi \, \kappa^2} =\frac{R_{\text{QPO}}}{\xi \, \kappa^2} ,
\end{equation}
where $\kappa$ and $\xi$ were taken to be 1.7, and 0.41, respectively \citep{shimura1995,kubota1998}.

For observations and segments in which there was only one feature in the kHz region (Obs-6 and Seg-3 of Obs-8), we assumed it to be the upper kHz QPO. The corresponding lower kHz QPO frequency was evaluated using the RPM relations for a mass and spin of $1.92 \, \text{M}_\odot$ and $0.145$, respectively \citep{Anand_2024}. The source distance and inclination angle were taken to be 5 kpc and 53 degrees, respectively \citep{Di_Salvo_2000,wang2019}.

In addition to the disk component, an additional blackbody component \verb|Bbodyrad| was added to the model to account for the emission from the boundary layer and/or the NS surface. \verb|Bbodyrad| has two parameters, namely the blackbody temperature (T$_{\text{bb}}$) and normalization. The \verb|Bbodyrad| normalization was not constrained; therefore, it was frozen to the value corresponding to the canonical NS radius of 10 km. The upper limit on the blackbody temperature was found to be $\sim$0.3--0.6 keV, and in some segments $\sim$0.8 keV where there was no or limited SXT exposure. Figures~\ref{fig:spectra}(a) and \ref{fig:spectra}(b) show representative energy spectra and fit residuals in the 0.7--20 keV range for Seg-1 of Obs-1 and Obs-3, respectively. The data points in red represent the SXT data, while the black points correspond to the LAXPC data. The best-fit spectral parameters of all observations and the corresponding segments have been tabulated in Table~\ref{tab:table_3} in Appendix~\ref{app:appA}.    

To demonstrate the statistical consistency of the spectral model adopted in this paper, we compare two alternative models. In the first model, the seed photons from the disk undergo Comptonization without any additional blackbody component, while in the second model, only the seed photons from the boundary layer and/or the NS surface are allowed to Comptonize. These alternative models are labeled as \verb|Model A: tbabs*(thcomp*diskbb + gaussian)| and \verb|Model B: tbabs(thcomp*bbodyrad + gaussian)|, respectively. It should be noted that the normalizations of both the \verb|diskbb| and \verb|bbodyrad| components were not well constrained during the fitting, as mentioned above. Therefore, we fixed them to values corresponding to the QPO radius for each segment and to the canonical NS radius of 10 km, respectively. We also fixed the width of the Gaussian component to 0.9 keV in those segments where it was not constrained in either model. 

\begin{figure*}
	\centering
	\includegraphics[width=0.9\textwidth]{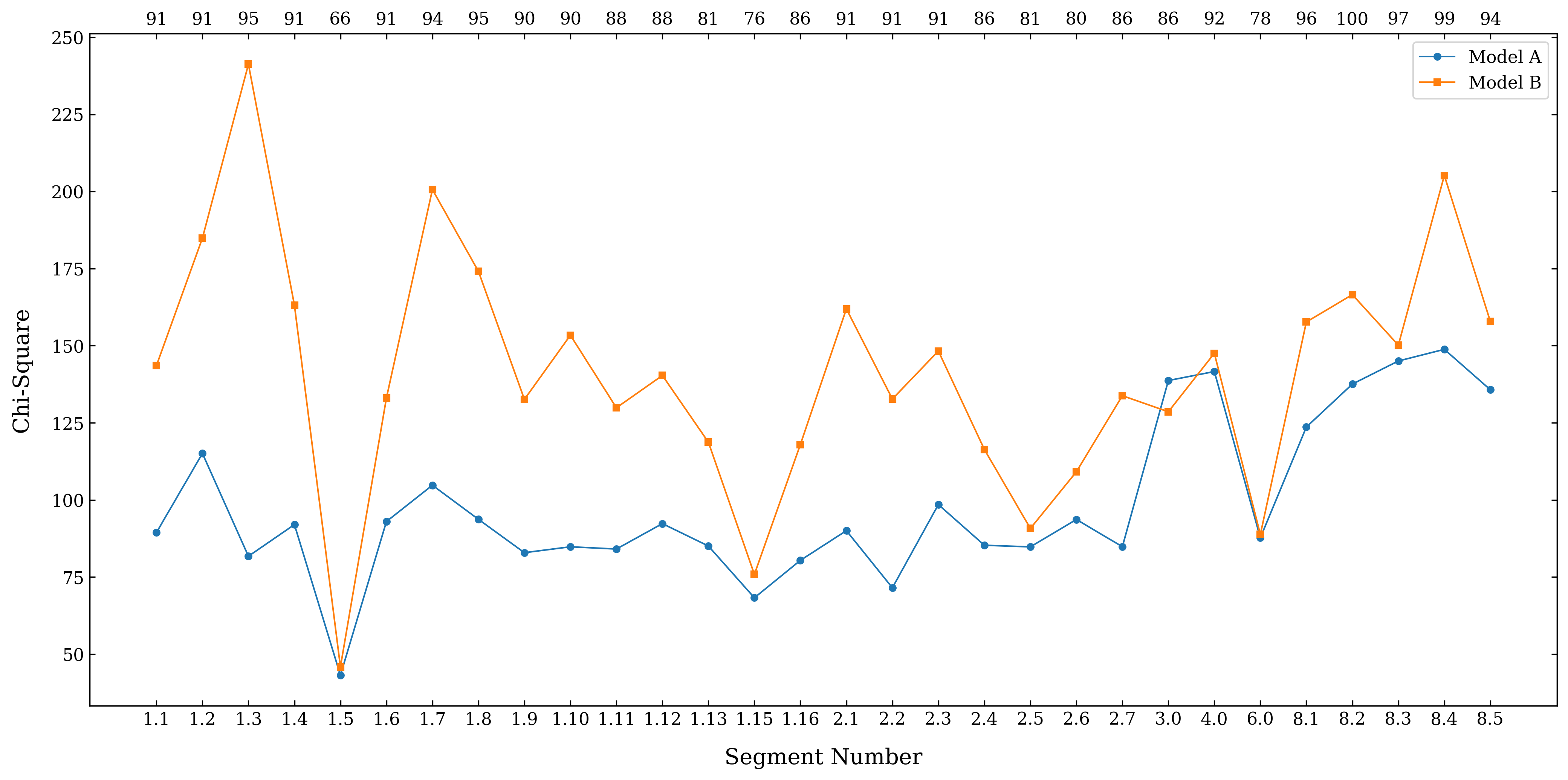}
	\caption{The figure shows the comparison of Chi-square between Model A: tbabs*(thcomp*diskbb + gaussian) and Model B: tbabs*(thcomp*bbodyrad + gaussian). The Y-axis shows Chi-square values and lower and upper X-axes show segment number and corresponding degrees of freedom, respectively. }
	\label{fig:chi_square}
\end{figure*}

\begin{figure*}
	\centering
	\begin{minipage}{0.45\textwidth}
		\includegraphics[angle=270,width=\textwidth]{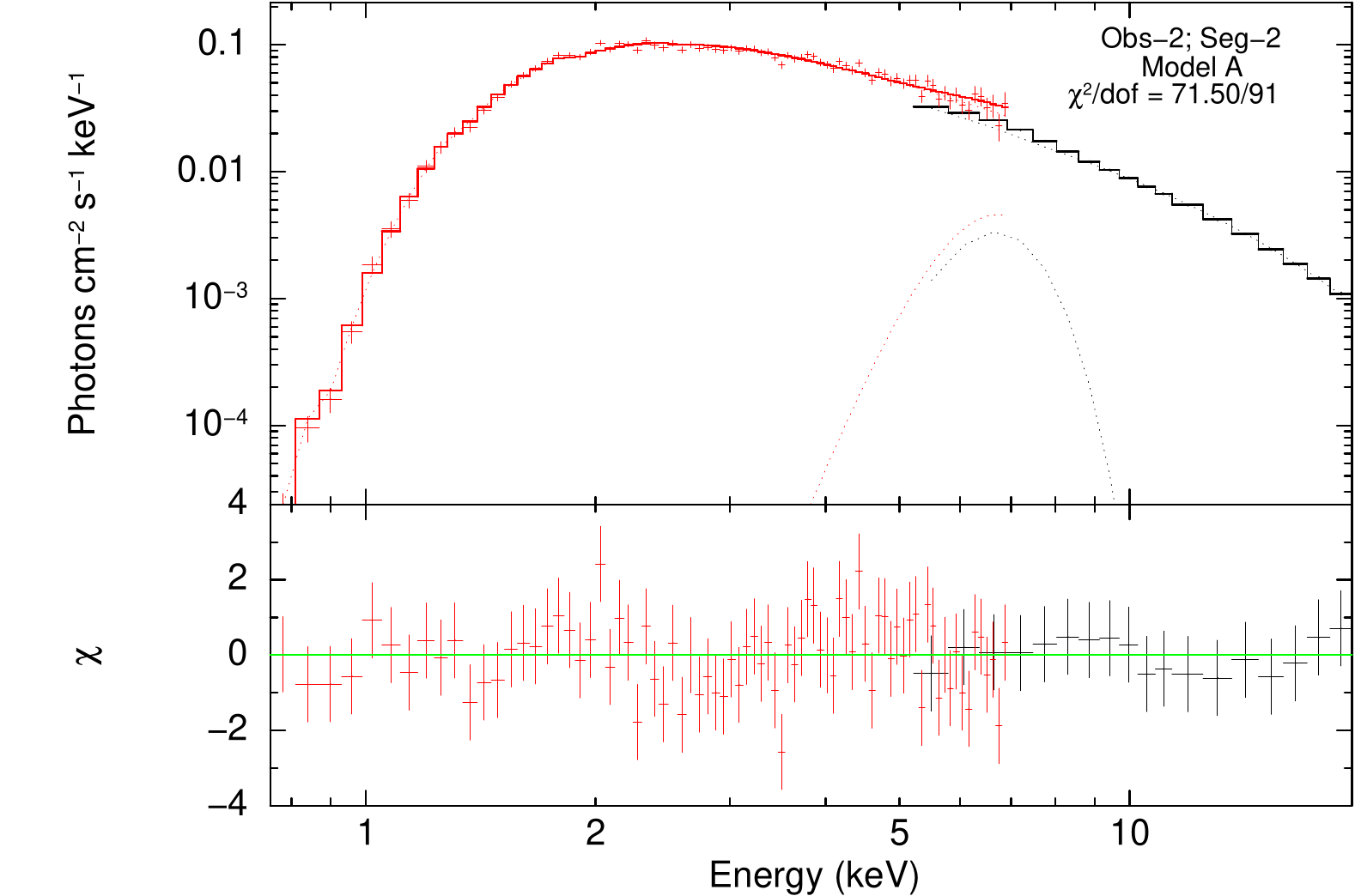}
		\centering {\small (a)}
	\end{minipage}
	\hspace{0.05\textwidth} 
	\begin{minipage}{0.45\textwidth}
		\includegraphics[angle=270,width=\textwidth]{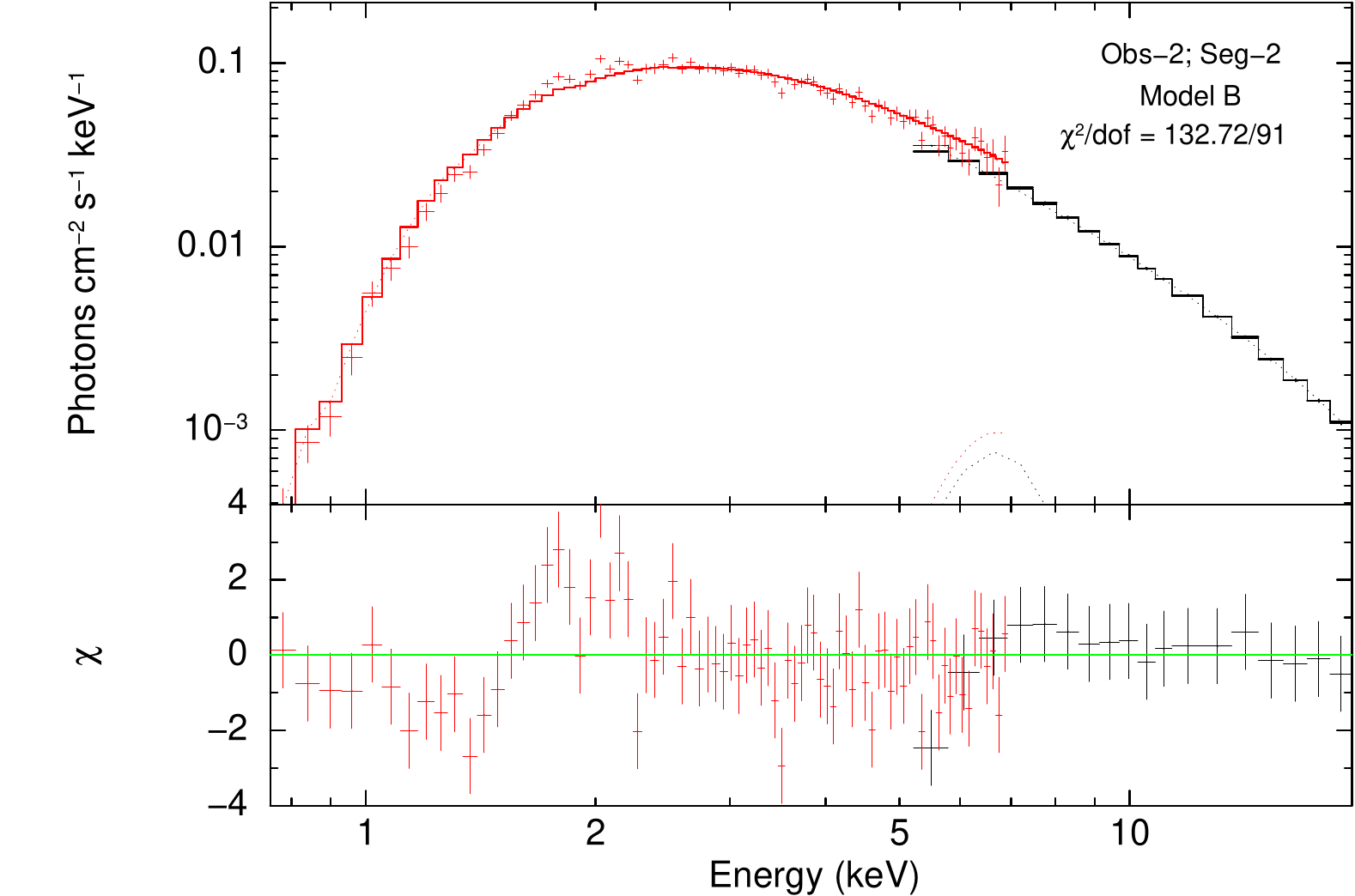}
		\centering {\small (b)}
	\end{minipage}

\vspace{8.5pt}
\begin{minipage}{0.45\textwidth}
\includegraphics[angle=270,width=\textwidth]{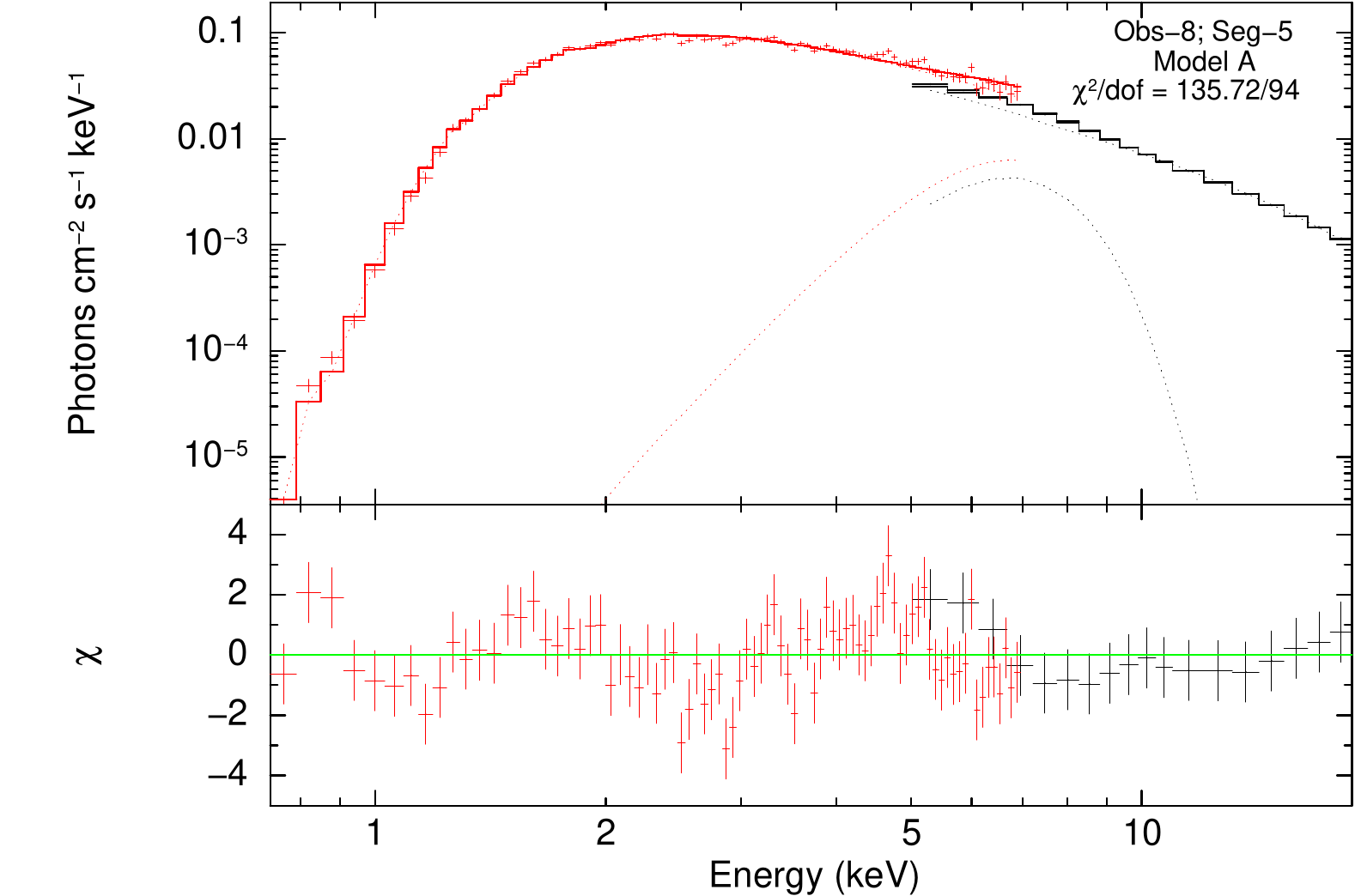}
\centering {\small (c)}
\end{minipage}
\hspace{0.05\textwidth} 
\begin{minipage}{0.45\textwidth}
\includegraphics[angle=270,width=\textwidth]{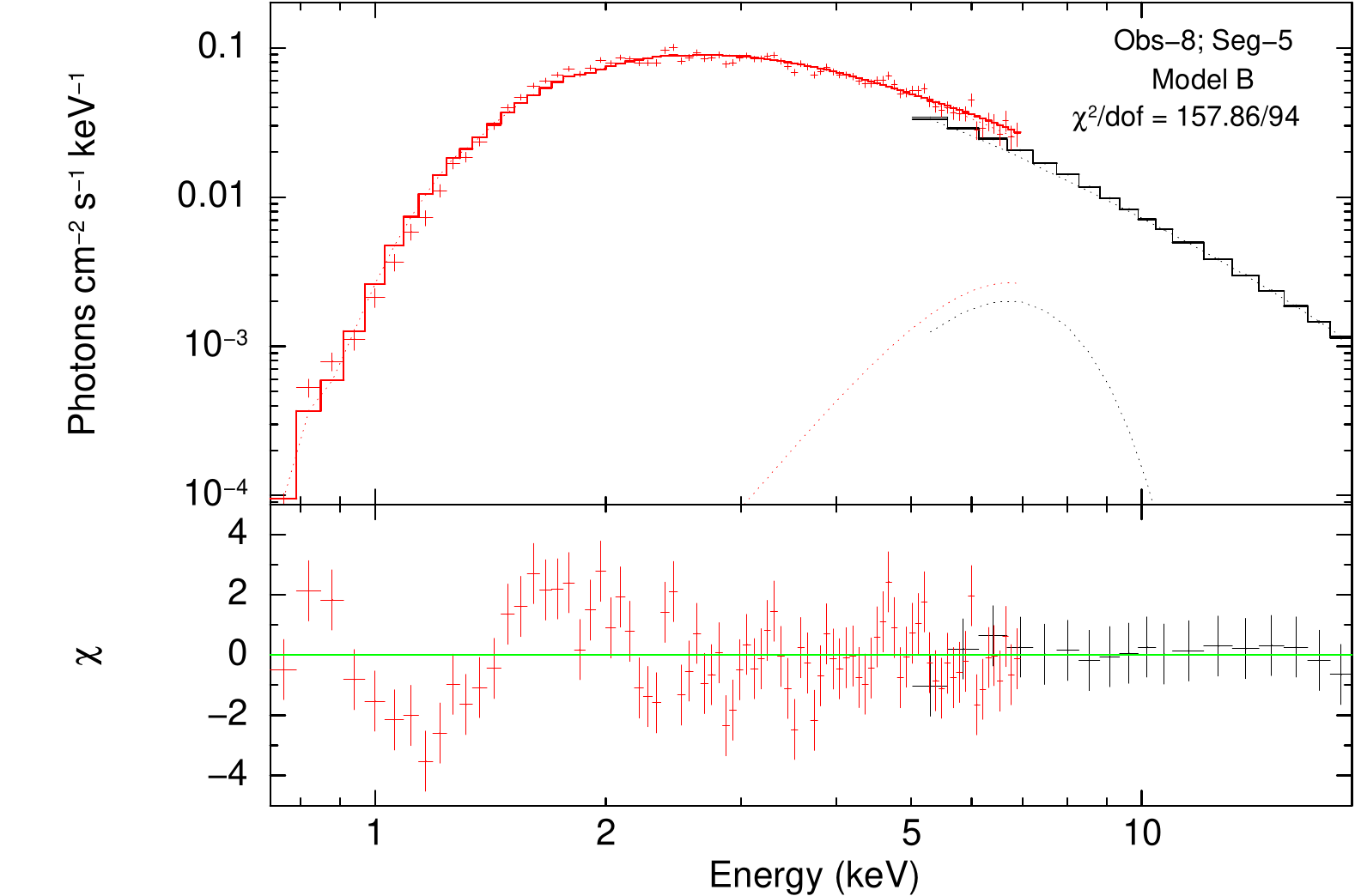}
\centering {\small (d)}
\end{minipage}

	\caption{The 0.7--20.0 keV SXT+LAXPC20 spectra for two representative segments are shown. The red data points correspond to the SXT spectra, while the black points represent the LAXPC spectra. The upper and lower panels show the spectra and the corresponding fit residuals for Seg-2 of Obs-2 and Seg-5 of Obs-8, respectively, fitted with Model A (left) and Model B (right).}
	\label{fig:spectra_2}
	
\end{figure*}

We compare the $\chi^2$ values of the two models only for those segments that have SXT exposure. Figure~\ref{fig:chi_square} shows that, for most segments, the $\chi^2$ value is significantly larger for Model B compared to Model A, except for Obs-3 (i.e., Seg-3.0), where $\chi_A^2 - \chi_B^2 \approx -10$. There are also a few segments where the difference in $\chi^2$ values ($\chi_A^2 - \chi_B^2$) is small (1–7). Two representative best-fit spectra of Seg-2 of Obs-2 and Seg-5 of Obs-8 (from both models), along with the fit residuals, are shown in Figure~\ref{fig:spectra_2}.

The $\chi^2$ values for Model A are significantly smaller than those for Model B in most of the segments, indicating that \verb|diskbb| is a dominant component. Therefore, Model A provides a better description of the data. In the final adopted model, Model A has been modified to include an additional blackbody component to account for the emission from the NS surface, thereby making the model physically plausible.

\section{Results}
Our analysis shows that the source underwent spectral transitions across the eight epochs. All the eight observations exhibit the presence of kHz QPOs, with frequencies ranging from $\sim$350 to 1180 Hz. Out of the 33 time segments analyzed across the eight observations, twin kHz QPOs were detected in 31 segments, while only a single kHz QPO was detected in the remaining two segments (see Table~\ref{tab:table_2}). By interpreting the origin of these QPOs within the framework of RPM, we infer that the inner disk radius varied from $\sim$16 to 27 km across the eight epochs.

\begin{figure*}
 	\centering
 	\begin{minipage}{0.45\textwidth}
 		\includegraphics[width=\textwidth]{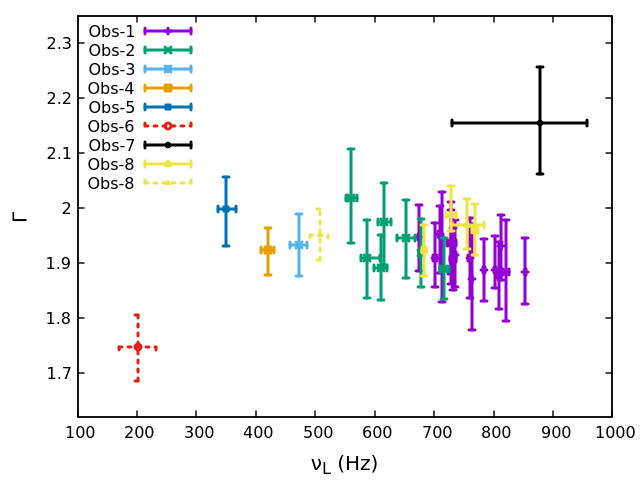}
 		\centering {\small (a)}
 	\end{minipage}
 	\begin{minipage}{0.45\textwidth}
 		\includegraphics[width=\textwidth]{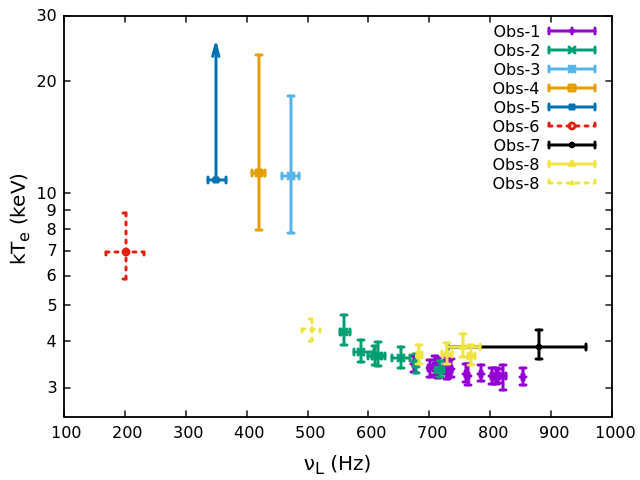}
 		\centering {\small (b)}
 	\end{minipage}
 	
 	\vspace{5pt}
 	
 	\begin{minipage}{0.45\textwidth}
 		\includegraphics[width=\textwidth]{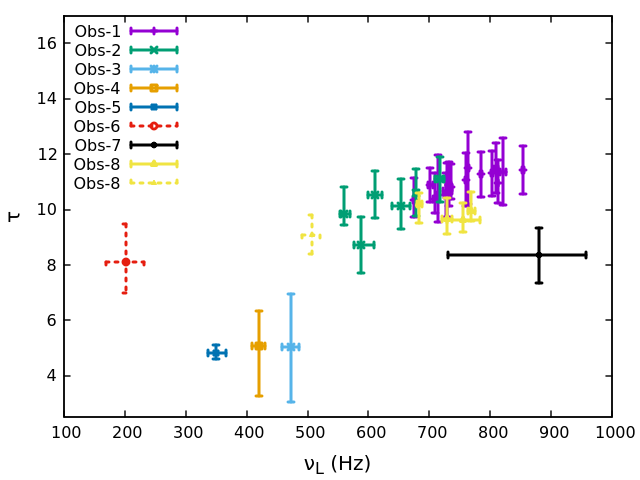}
 		\centering {\small (c)}
 	\end{minipage}
 	\begin{minipage}{0.45\textwidth}
 		\includegraphics[width=\textwidth]{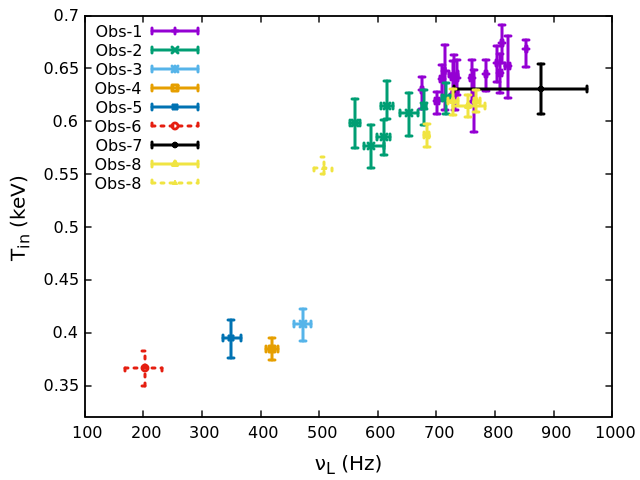}
 		\centering {\small (d)}
 	\end{minipage}

 	\caption{The figure shows the correlation between the lower kHz QPO frequency and (a) the spectral index, (b) electron temperature, (c) Thomson optical depth, and (d) inner disk temperature. Data points connected by dashed lines correspond to segments in which only a single kHz QPO was detected. The data point represented by an arrow indicates that the upper limit of the parameter is unconstrained in panel (b). Since kT$_e$ for Obs-5 was unconstrained, the corresponding $\tau$ was evaluated by freezing kT$_e$ at 11 keV.}
 	\label{fig:corr1}
 \end{figure*}

\begin{figure*}
	\centering
	\begin{minipage}{0.45\textwidth}
		\includegraphics[width=\textwidth]{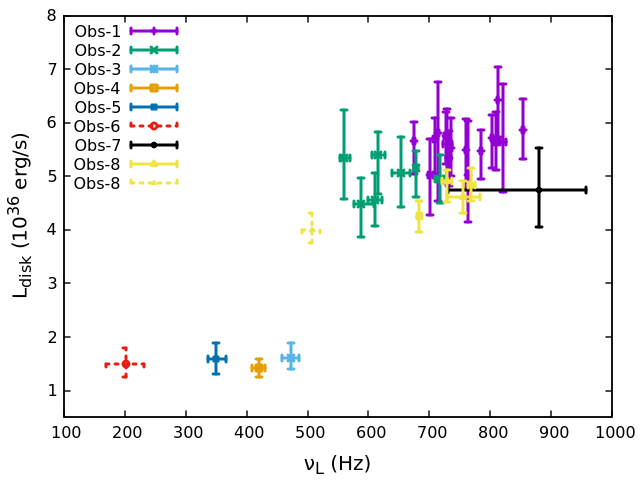}
		\centering {\small (a)}
	\end{minipage}
	\begin{minipage}{0.45\textwidth}
		\includegraphics[width=\textwidth]{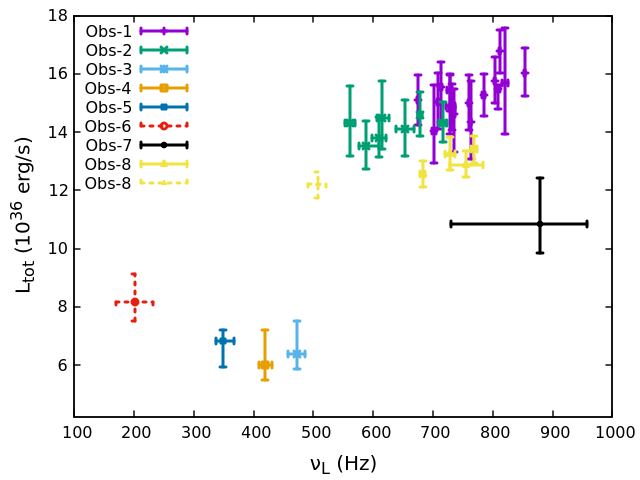}
		\centering {\small (b)}
	\end{minipage}
	
	\vspace{5pt}
	
	\begin{minipage}{0.45\textwidth}
		\includegraphics[width=\textwidth]{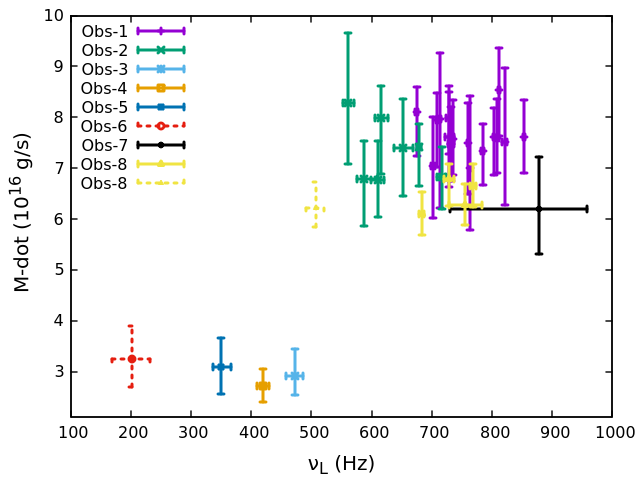}
		\centering {\small (c)}
	\end{minipage}
	\begin{minipage}{0.45\textwidth}
		\includegraphics[width=\textwidth]{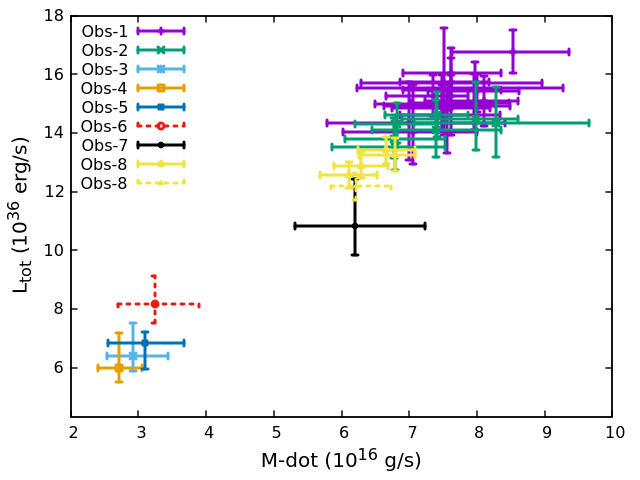}
		\centering {\small (d)}
	\end{minipage}

	\caption{The correlations of the lower kHz QPO frequency with the disk luminosity (a), total luminosity (b), and mass accretion rate (c) are shown. Figure~\ref{fig:corr1}(d) shows the correlation between the mass accretion rate and total luminosity. }
	\label{fig:corr2}
\end{figure*}

The time-resolved spectral analysis of the persistent emission shows correlations between various spectral parameters and the lower kHz QPO frequency ($\nu_{\text{L}}$). In the correlation plots presented below (Figure~\ref{fig:corr1} \& Figure~\ref{fig:corr2}), two data points are indicated by dashed lines; these correspond to Obs-6 and Seg-3 of Obs-8, where only a single kHz QPO was detected. These were assumed to be the upper kHz QPOs, and the corresponding lower kHz QPO frequencies were estimated using RPM as described in \cite{Anand_2024}.

Figure~\ref{fig:corr1} shows the correlations between the spectral parameters, derived from spectral fitting, and $\nu_{\text{L}}$. The photon index remains approximately constant at $\sim$ 2 across most observations, with the exception of Obs-6, which exhibits a slightly lower value of $\sim1.7$. Overall, $\Gamma$ shows no clear correlation with $\nu_{\text{L}}$. The Comptonizing region appears to be hotter ($\gtrsim$10 keV) when $\nu_{\text{L}}$ is below 500 Hz, kT$_e$ decreases almost linearly from $\sim$5 to 3 keV as $\nu_{\text{L}}$ increases beyond 500 Hz.

Figures~\ref{fig:corr1}(c) and \ref{fig:corr1}(d) show the correlations of $\nu_{\text{L}}$ with $\tau$ and T$_{\text{in}}$, respectively. The optical depth is $\sim$5 below 500 Hz and increases from $\sim$8 to 12, showing a strong positive correlation with $\nu_{\text{L}}$ above 500 Hz. On the other hand, T$_{\text{in}}$ exhibits a positive correlation with $\nu_{\text{L}}$ both below and above 500 Hz, although the slopes differ between the two regions as shown in Figures~\ref{fig:corr1}(d).

We also calculated the unabsorbed disk and total fluxes for each segment using the convolution model \verb|cflux|. These fluxes were then converted to luminosities using the known source distance of 5 kpc. Figures~\ref{fig:corr2}(a) and \ref{fig:corr2}(b) show the variations of the disk luminosity (L$_{\text{disk}}$) and the unabsorbed total luminosity (L$_{\text{tot}}$) with $\nu_{\text{L}}$. As apparent from the figures, L$_{\text{disk}}$ remains nearly constant at $\sim1.5 \times 10^{36} \, \text{erg} \, \text{s}^{-1}$ for frequencies below 500 Hz, and shows a moderate positive correlation above 500 Hz, increasing from $4$–$6 \times 10^{36} \, \text{erg} \, \text{s}^{-1}$. The total luminosity  also shows a positive correlation above 500 Hz, ranging from $12$–$17 \times 10^{36} \, \text{erg} \, \text{s}^{-1}$. However, below 500 Hz, L$_{\text{tot}}$ is slightly higher in Obs-6 compared to Obs-3, Obs-4, and Obs-5.

We computed the mass accretion rate ($\dot{\text{M}}$) from the disk luminosity using the following relation:
\begin{equation}
\text{L}_{\text{disk}} = \frac{1}{2}\, \frac{GM\dot{M}}{R_{\text{in}}},
\end{equation}
here, M is the mass of the central object, and R$_{\text{in}}$ is the inner disk radius. Using the known mass of the NS (1.92 M$_\odot$ in our case) and estimating R$_{\text{in}}$ from the upper kHz QPO frequency, we calculated the mass accretion rate ($\dot{\text{M}}$) for all the segments. Figure~\ref{fig:corr2}(c) shows the variation of $\dot{\text{M}}$ with $\nu_{\text{L}}$. The data clearly reveal two accretion regimes, namely AR1 and AR2, with accretion rates of $\sim3 \times 10^{16} \, \text{g} \, \text{s}^{-1}$ for $\nu_{\text{L}} < 500$ Hz, and around $7 \times 10^{16} \, \text{g} \, \text{s}^{-1}$ for $\nu_{\text{L}} > 500$ Hz, respectively. Within each state, ($\dot{\text{M}}$ remains nearly constant, showing no dependence on the QPO frequency. It is evident that the accretion rate is about 2.3 times higher in the soft state (Obs-1, 2, 7, and 8) compared to the hard state (Obs-3, 4, 5, and 6).

\begin{figure}
 	
 	\includegraphics[width=\columnwidth]{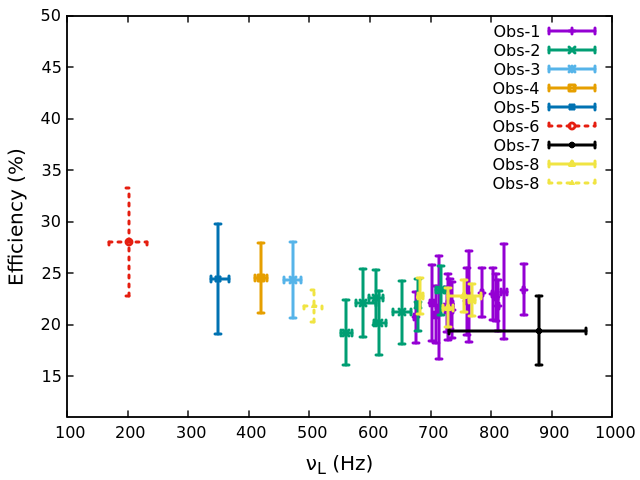}
 	\caption{The figure shows the accretion efficiency as a function of the lower kHz QPO frequency.}
 	\label{fig:effi}
 	
 \end{figure}

Figure~\ref{fig:corr2}(d) presents the correlation between total luminosity and mass accretion rate, with a Spearman correlation coefficient ($r_{\text{s}}$) of 0.86 and a p-value of $ 9.1\times 10^{-11}$. Although $\dot{\text{M}}$ decreased by the factor of 2.3, it still maintained a strong positive correlation with the total luminosity, as expected. We also estimated the accretion efficiency for all segments, defined as $\eta = L_{\text{tot}} / \dot{M}c^2$. Figure~\ref{fig:effi} shows that the efficiency is $\sim$22 \% and remains nearly constant across all observations within the 90\% confidence interval. A slightly higher efficiency can be seen below $\sim$500 Hz, particularly during Obs-6.

\section{Summary \& Discussion}
We conducted a comprehensive spectro-timing analysis of all available AstroSat archival data from 2016 to 2019 for the NS-LMXB 4U 1728–34. We find that:
\begin{itemize}
	\item The source undergoes spectral transitions, with four distinct regions identified in HID. These regions represent four different spectral states exhibited by the source during its spectral evolution. 
	
	\item The four spectral states correspond to two accretion regimes: $\sim3 \times 10^{16} \, \text{g} \, \text{s}^{-1}$ (AR1) for Obs-3, 4, 5, and 6, and about $7 \times 10^{16} \, \text{g} \, \text{s}^{-1}$ (AR2) for Obs-1, 2, 7, and 8. Within each accretion regime, the accretion rate remains nearly constant.
	
	\item We detected kHz QPOs in all eight observations, with frequencies ranging from $\sim$350 to 1180 Hz, although some QPO detections were less significant ($<$3$\sigma$). The detected upper kHz QPOs correspond to inner disk radii ranging from $\sim 27$ to $\sim 16$ km.
	
	\item kT$_{\text{e}}$ remains $\sim$10 keV in AR1. Beyond 500 Hz (corresponding to an inner disk radius of $\sim$19 km), i.e., in AR2, kT$_{\text{e}}$ decreases linearly, showing a strong anti-correlation with $\nu_{\text{L}}$ ($r_{\text{s}} = -0.78$, p-value = $1.14 \times 10^{-6}$). While, $\tau$ increases from $\sim 8$ to 12 in AR2, showing a positive correlation with $\nu_{\text{L}}$ ($r_{\text{s}} = 0.71$, p-value = $1.96 \times 10^{-5}$). These trends are consistent with results obtained for 4U 1636–536 from the RXTE observations \citep{Ribeiro2017}.
	
\end{itemize}

Since the lower and upper kHz QPOs are known to be tightly correlated \citep{Anand_2024}, we focus only on the lower kHz QPO when exploring the QPO correlations with spectral parameters. We find no systematic dependence of the lower kHz QPO frequency on the spectral index, which contrasts with results reported for 4U 1636–536 by \cite{Ribeiro2017}. \cite{Ribeiro2017} observed an anti-correlation between $\Gamma$ and $\nu_{\text{L}}$ in the range of $\sim$ 500--900 Hz. However, when considering only the data from Obs-1, Obs-2, and Obs-8, that is in the frequency range of $\sim$500 to 900 Hz, a weak negative correlation emerges, with $r_{\text{s}}$ of $-0.5$ and a p-value of $6.2 \times 10^{-3}$. This anti-correlation is in contrast to black hole systems, where Type-C QPO frequency is typically positively correlated with $\Gamma$ \citep{vignarca2003}.

It should be noted that in Obs-6 and Seg-3 of Obs-8, only a single kHz QPO was detected, represented by dashed lines in the correlation plots. It is not possible to definitively classify these as either lower or upper QPOs. The QPO detected in Seg-3 of Obs-8 has a frequency of $\sim$910 Hz, close to the upper kHz QPOs identified in other segments of the same observation. Therefore, we treat this QPO as an upper kHz QPO. However, the QPO detected in Obs-6 is less clear. If interpreted as a lower kHz QPO, the estimated accretion efficiency would be $\sim$ 35\%, which is significantly higher than expected for NS systems. Thus, we treat this as the upper kHz QPO, which yields an efficiency consistent with the theoretically expected value.

As shown in the correlation plots, the spectral parameters for Seg-14 of Obs-1, Obs-5, and Obs-7 are either poorly constrained or have large uncertainties. This is due to the absence of SXT data for these segments. Given the moderate energy resolution of the LAXPC and limited photon statistics, the spectral parameters could not be tightly constrained from LAXPC data alone.

In NS systems with a weak magnetic field ($< 10^8$ G), the accretion disk may extend up to the surface of the star. In such a scenario, there will also be emission from the boundary layer. The luminosity from the boundary layer can be expressed as
\begin{equation}
\text{L}_{\text{BL}}=\frac{1}{2}\frac{GM\dot{M}}{R_*}\left(1-\frac{\nu_s}{\nu_k}\right)^2, 
\end{equation}
here, $\nu_s$ is the spin frequency, $R_*$ is the radius of the NS, and $\nu_k$ is the Keplerian frequency at the stellar surface. Thus, the total luminosity is the sum of the disk luminosity and the boundary layer luminosity \citep{Gilfanov_2014}. For a NS mass of 1.92 M$\odot$ and assuming that the accretion disk extends to the NS surface (with $R* = 10$ km), the theoretical value of the efficiency is estimated to be $\sim$24 \%.  

The kHz QPOs in NS systems are known to correlate with the spectral state of the source, as discussed above. In several sources, the detection of kHz QPOs, particularly the lower kHz QPO, depends strongly on the position of the source in CD, and typically occurs during transitions from the hard to the soft state \citep{Mendez_1999,Mendez_2001,Zhang_2017}.

\cite{Zhang_2017} performed a detailed study of the relation between the presence of kHz QPOs and the spectral evolution of 4U 1636--536, showing that during the transition from the hard to the soft state, i.e., when the lower kHz QPO is present, $\Gamma$ drops abruptly, while $kT_e$ decreases gradually and $\tau$ remains relatively high ($\sim$5--25). Similarly, \cite{Ribeiro2017} investigated the correlations between the kHz QPO frequency and spectral parameters and found that $\nu_{\mathrm{L}}$ is anti-correlated with $\Gamma$ and $kT_e$, and positively correlated with $\tau$.

For the first time, we have carried out a similar analysis for 4U 1728--34, as previously done by \cite{Ribeiro2017} for 4U 1636--536, by exploiting the fine time resolution of LAXPC, complemented by a better spectral resolution of SXT, which together enabled us to investigate spectro-timing correlations. Our results are largely consistent with those of \cite{Zhang_2017} and \cite{Ribeiro2017}. In addition, we extend this analysis by demonstrating correlations between $\nu_{\mathrm{L}}$ and T$_{\text{in}}$, the luminosity, and $\dot{\text{M}}$. We find evidence for a critical frequency at $\nu_{\mathrm{L}} \sim 500$~Hz, at which the source undergoes a sudden spectral transition between the hard and soft states, while within each state the accretion rate remains approximately constant. Finally, our results indicate that the radiative efficiency is nearly constant at $\sim$22 \%, and matches with the theoretically expected value and shows no significant dependence on the spectral state of the source.

To summarize, our results demonstrate a clear dependence of the kHz QPO frequency on spectral parameters, emphasizing the existence of a critical QPO frequency. This critical frequency is influenced by the source’s spectral transition, and the nature of the correlations may differ above and below this threshold. To gain a deeper understanding of the accretion dynamics in NS–LMXBs, it is therefore essential to explore the relationship between their spectral and timing characteristics in a larger sample. While only a limited number of studies have investigated the correlations between kHz QPOs and spectral parameters, similar analyses should be carried out for other sources that exhibit kHz QPOs using the RXTE and AstroSat archival data. We also hope that the joint campaigns with NICER and NuSTAR will significantly enhance our understanding of these complex phenomena in such systems.

\section{Acknowledgments}

We express our gratitude to the anonymous reviewer for the insightful remarks and constructive suggestions. This work utilized data from the LAXPC and SXT instruments onboard AstroSat, an observatory launched and operated by ISRO. We acknowledge the support of the SXT and LAXPC Payload Operations Centres at TIFR, as well as the AstroSat Science Support Cell at IUCAA, for making the data and software available. The data analysis was performed using LAXPCsoftware, the SXT pipeline, and HeaSoft-6.30.1 of NASA’s High Energy Astrophysics Science Archive Research Center. K.A. expresses his sincere gratitude to IUCAA for providing funding to support multiple visits throughout the course of this work.

\appendix

\section{Spectro-timing study}
\label{app:appA}
The power density spectra (PDS) were generated from the LAXPC data for each individual segment. For the first three observations, data from all three LAXPC units were used, while for the remaining observations, only LAXPC20 was used. The PDS were fitted with multiple Lorentzian components, along with a power-law with zero spectral index to account for the Poisson noise residual. Table~\ref{tab:table_2} presents only the best-fit parameters of the high-frequency Lorentzians, along with their significance, the RMS values, and the LAXPC exposure time for each individual segment.\\

We fitted the combined SXT and LAXPC20 spectra in the energy ranges of 0.7–7.0 keV and 5–20 keV, respectively, for each individual segment using the model described in Section~\ref{sec:sec3.2}. For the segments with no SXT exposure, only the LAXPC20 spectra were fitted. The best-fit spectral parameters for each individual segment are listed in Table~\ref{tab:table_3}.  

\begin{longtable}{ccccccc}
	\caption{The best-fit parameters of the kHz QPOs or HF Lorentzians for each segment. The errors are within the 90\% confidence interval.}
	\label{tab:table_2} \\
	
	\hline
	Obs\#.Seg\# & Exposure Time (s) & \multicolumn{4}{c}{kHz QPO(s) or HF Lorentzian} & $\chi^2/\text{dof}$ \\
	\cline{3-6}
	&  & $\nu$ (Hz) & $\Delta \nu$ (Hz) & $s$ & RMS (\%) & \\
	\hline 
	\endfirsthead
	
	\hline
	\multicolumn{7}{c}{\textit{Table \thetable\ (continued)}} \\
	\hline
	Obs\#.Seg\# & Exposure Time (s) & \multicolumn{4}{c}{kHz QPO(s) or HF Lorentzian} & $\chi^2/\text{dof}$ \\
	\cline{3-6}
	&  & $\nu$ (Hz) & $\Delta \nu$ & $s$ & RMS (\%) &   \\
	\hline
	\endhead
	
	\hline \multicolumn{7}{r}{\textit{Continued on the next page}}  \\ 
	\endfoot  
	\hline
	\endlastfoot
	
	1.1 & 2902 & $853.4_{-1.3}^{+1.1}$ & $16.9_{-2.2}^{+3.4}$ & 20.68 & $7.4_{-0.3}^{+0.4}$ & 60.30/50  \vspace{3pt} \\
	& & $1186_{-72}^{+85}$ & $> 162.52$ & 3.45 & $6.5_{-1.4}^{+0.9}$ &  \vspace{3pt}  \\
	
	\hline  
    1.2 & 3327 & $802.9_{-1.6}^{+0.7}$ & $17.2_{-3.2}^{+2.8}$ & 13.17 & $7.6_{-0.4}^{+0.4}$ & 82.54/65 \vspace{3pt} \\
    & & $1142_{-29}^{+47}$ & $> 23.42$ & 2.39 & $4.6_{-1.5}^{+2.4}$ & \vspace{3pt} \\

	\hline 
	1.3 & 7204 & $811.9_{-1.6}^{+0.8}$ & $26.7_{-3.3}^{+3.2}$ & 18.26
	& $7.8_{-0.3}^{+0.3}$ & 121.93/107  \vspace{3pt} \\
	& & $1145_{-18}^{+17}$ & $82_{-29}^{+42}$ & 4.59 &
	$4.7_{-0.8}^{+0.9}$ &  \vspace{3pt} \\
	
	\hline
	1.4 & 5714 & $808.5_{-3.5}^{+3.3}$ & $56.1_{-7.8}^{+8.4}$ & 14.03
	& $8.4_{-0.5}^{+0.5}$ & 129.13/96  \vspace{3pt} \\
	& & $1126_{-27}^{+29}$ & $156_{-65}^{+120}$ & 3.89 &
	$6.1_{-1.2}^{+1.8}$ &  \vspace{3pt} \\
	
	\hline 
1.5 & 4206 & $821.2_{-5.0}^{+4.5}$ & $68_{-10}^{+11}$ & 13.08 & $9.0_{-0.6}^{+0.6}$ & 91.53/69 \vspace{3pt} \\
& & $1141_{-10}^{+22}$ & $< 102.29$ & 2.19 & $2.7_{-0.9}^{+2.3}$ & \vspace{3pt} \\
	
	\hline
	1.6 & 3890 & $735.3_{-2.1}^{+2.0}$ & $30.1_{-5.2}^{+6.1}$ & 12.15
	& $7.5_{-0.5}^{+0.5}$ & 113.94/101  \vspace{5pt} \\
	& & $1100_{-18}^{+20}$ & $153_{-60}^{+100}$  & 4.38 &
	$7.9_{-1.4}^{+2.2}$ &   \vspace{3pt} \\
	
	\hline
1.7 & 3265 & $727.8_{-3.8}^{+2.6}$ & $32.2_{-8.7}^{+8.0}$ & 8.02
& $7.0_{-0.7}^{+0.6}$ & 86.59/84  \vspace{3pt} \\
& & $1075_{-20}^{+15}$ & $111_{-50}^{+73}$  & 4.05 &
$7.0_{-1.3}^{+1.6}$ &   \vspace{3pt} \\
	
	\hline
	1.8 & 4156 & $784.7_{-2.2}^{+1.6}$ & $30.0_{-5.9}^{+6.3}$ & 11.36
	& $7.6_{-0.5}^{+0.5}$ & 75.93/74   \vspace{3pt} \\
	& & $1129_{-25}^{+25}$ & $133_{-52}^{+99}$  & 3.91 &
	$6.3_{-1.2}^{+1.9}$ &   \vspace{3pt} \\
	
	\hline
	1.9 & 3442 & $731.4_{-4.7}^{+4.7}$ & $58.2_{-11}^{+13}$ & 10.08 & $8.4_{-0.7}^{+0.7}$ & 81.16/84 \vspace{3pt} \\
	& & $1063_{-17}^{+20}$ & $159_{-61}^{+97}$ & 4.64 & $8.7_{-1.5}^{+2.2}$ & \vspace{3pt} \\
	
	\hline
	1.10 & 3619 & $675.0_{-2.2}^{+1.8}$ & $20.6_{-5.5}^{+7.1}$ & 8.34
	& $6.0_{-0.6}^{+0.6}$ & 83.20/98   \vspace{3pt} \\
	& & $1030_{-11}^{+11}$ & $90_{-30}^{+46}$  & 5.68 &
	$7.5_{-1.0}^{+1.3}$ &   \vspace{3pt} \\
	
	\hline
	1.11 & 2773 & $709.2_{-3.0}^{+2.7}$ & $28.7_{-5.8}^{+7.0}$ & 9.67
	& $7.2_{-0.6}^{+0.6}$ & 85.43/69  \vspace{3pt} \\
	& & $1071_{-25}^{+45}$ & $140_{-74}^{+285}$  & 3.11 &
	$7.2_{-1.7}^{+6.1}$ & \vspace{3pt} \\
	
	\hline 
	1.12 & 3057 & $759.9_{-2.2}^{+1.2}$ & $22.7_{-3.5}^{+3.7}$ &
	13.29 & $7.9_{-0.5}^{+0.5}$ & 87.11/62   \vspace{3pt} \\
	& & $1106_{-13}^{+15}$ & $71_{-30}^{+55}$  & 4.14 &
	$6.0_{-1.1}^{+1.6}$ &   \vspace{3pt} \\
	
	\hline
    1.13 & 2025 & $714.0_{-2.0}^{+1.1}$ & $17.3_{-4.3}^{+4.9}$ & 9.03
    & $7.2_{-0.6}^{+0.6}$ & 54.60/58  \vspace{3pt} \\
    & & $1095_{-21}^{+17}$ & $105_{-48}^{+99}$  & 3.79 &
    $7.5_{-1.5}^{+2.6}$ & \vspace{3pt} \\
	
	\hline
	 1.14 & 2210 & $763.1_{-0.6}^{+0.8}$ & $11.4_{-1.2}^{+1.3}$ &
	 13.92 & $7.4_{-0.4}^{+0.4}$ & 55.52/52 \vspace{3pt} \\
	 & & $1071_{-13}^{+14}$ & $67_{-29}^{+50}$ & 4.35 &
	 $6.7_{-1.1}^{+1.6}$ & \vspace{3pt} \\
	\hline
	1.15 & 1931 & $701.6_{-3.4}^{+3.5}$ & $30_{-9}^{+13}$ &
	6.79 & $7.0_{-0.8}^{+1.0}$ & 74.43/65 \vspace{3pt} \\
	& & $1058_{-19}^{+34}$ & $103_{-59}^{+120}$ & 3.26 &
	$7.8_{-1.8}^{+3.2}$ & \vspace{3pt} \\
	
	\hline
	1.16 & 2373 & $729.0_{-6.9}^{+8.4}$ & $58_{-26}^{+34}$ &
	4.97 & $7.3_{-1.2}^{+1.4}$ & 65.23/66 \vspace{3pt} \\
	& & $1110_{-29}^{+29}$ & $168_{-83}^{+170}$ & 3.36 &
	$8.1_{-1.8}^{+3.1}$ & \vspace{3pt} \\
	
\hline
2.1 & 3045 & $610_{-11}^{+11}$ & $51_{-21}^{+40}$ &
4.09 & $5.7_{-1.1}^{+1.5}$ & 96.67/94 \vspace{3pt} \\
& & $977.2_{-7.1}^{+7.2}$ & $77.2_{-18}^{+18}$ & 7.29 &
$9.6_{-1.0}^{+1.2}$ & \vspace{3pt} \\
\hline
2.2 & 2128 & $678.4_{-3.2}^{+3.3}$ & $25_{-8}^{+11}$ & 6.71
& $6.7_{-0.8}^{+0.9}$ & 61.65/66 \vspace{3pt} \\
& & $1024_{-19}^{+25}$ & $110_{-65}^{+90}$ & 3.33 &
$7.9_{-1.9}^{+2.3}$ & \vspace{3pt} \\
\hline
2.3 & 2347 & $716.6_{-7.3}^{+7.7}$ & $44_{-27}^{+22}$
& 3.20 & $7.5_{-1.6}^{+1.2}$ & 93.09/62 \vspace{3pt} \\
& & $1091_{-23}^{+25}$ & $90_{-42}^{+70}$ &
3.06 & $6.4_{-1.6}^{+1.9}$ & \vspace{3pt} \\
\hline
2.4 & 2685 & $652_{-15}^{+16}$ &
$90_{-32}^{+50}$ & 5.01 & $7.6_{-1.2}^{+1.5}$ &
99.49/102 \vspace{3pt} \\
& & $1004.4_{-9.1}^{+9.2}$ & $81_{-26}^{+36}$ & 6.25
& $8.8_{-1.1}^{+1.3}$ & \vspace{3pt} \\
\hline
2.5 & 1763 & $616_{-9}^{+11}$ &
$35_{-19}^{+37}$ & 3.34 & $5.4_{-1.3}^{+1.7}$ &
76.43/67 \vspace{3pt} \\
& & $986_{-14}^{+14}$ & $129_{-42}^{+66}$ &
5.45 & $11.0_{-1.6}^{+2.2}$ & \vspace{3pt} \\
\hline
2.6 & 4385 & $561_{-8.1}^{+9.4}$ & $20_{-16}^{+57}$
& 2.40 & $3.4_{-1.1}^{+1.8}$ & 153.32/138 \vspace{3pt} \\
& & $923.2_{-7.6}^{+7.7}$ & $101_{-22}^{+29}$ & 8.63
& $10.3_{-1.0}^{+1.1}$ & \vspace{3pt} \\
\hline
2.7 & 2836 & $588_{-11}^{+22}$ & $30_{-25}^{+88}$ &
2.34 & $4.0_{-1.3}^{+2.4}$ & 81.96/101 \vspace{3pt} \\
& & $950.7_{-8.4}^{+8.8}$ & $89_{-28}^{+39}$ & 6.35 &
$9.7_{-1.2}^{+1.5}$ & \vspace{3pt} \\
\hline
3.0 & 7637 & $472_{-15}^{+14}$ & $52_{-23}^{+41}$ &
3.70 & $7.9_{-1.7}^{+2.2}$ & 166.96/217 \vspace{3pt} \\
& & $732_{-30}^{+15}$ & $>18.74$ & 2.83 &
$6.6_{-1.8}^{+3.2}$ & \vspace{3pt} \\
\hline
4.0 & 10220 & $420_{-10}^{+10}$ & $32_{-15}^{+31}$ &
3.31 & $9.6_{-2.5}^{+2.8}$ & 93.17/103 \vspace{3pt} \\
& & $678_{-68}^{+43}$ & $>6.96$ & 2.11 &
$10.4_{-4.1}^{+3.2}$ & \vspace{3pt} \\
\hline
5.0 & 38632 & $350_{-6}^{+18}$ & $>18.05$ & 3.61 &
$5.1_{-1.1}^{+1.0}$ & 229.99/207 \vspace{3pt} \\
& & $665_{-38}^{+31}$ & $>18.46$ & 1.83 &
$6.5_{-2.3}^{+1.5}$ & \vspace{3pt} \\
\hline
6.0 & 7449 & $566_{-53}^{+44}$ & $186_{-174}^{+274}$ &
2.56 & $10.3_{-4.4}^{+5.2}$ & 72.43/75 \vspace{3pt} \\
\hline
7.0 & 84100 & $879_{-150}^{+78}$ & $<78^*$ & 1.59 &
$3.0_{-1.4}^{+1.1}$ & 192.70/182 \vspace{3pt} \\
& & $1173_{-41}^{+52}$ & $>28.39$ & 2.49 &
$4.9_{-1.7}^{+1.1}$ & \vspace{3pt} \\
\hline
8.1 & 16761 & $683.1_{-4.3}^{+3.9}$ & $17_{-11}^{+21}$ &
3.72 & $4.8_{-1.0}^{+1.4}$ & 90.68/95 \vspace{3pt} \\
& & $1028_{-21}^{+22}$ & $154_{-62}^{+105}$ & 4.07
& $9.9_{-1.8}^{+2.5}$ & \vspace{3pt} \\
\hline
8.2 & 17599 & $768.2_{-4.6}^{+5.9}$ & $24_{-14}^{+27}$ &
3.72 & $5.4_{-1.0}^{+1.4}$ & 97.45/79 \vspace{3pt} \\
& & $1091.5_{-5.9}^{+5.9}$ & $20_{-20}^{+34}$ & 3.17 &
$4.6_{-1.1}^{+1.7}$ & \vspace{3pt} \\
\hline
8.3 & 16756 & $914_{-15}^{+12}$ & $110_{-38}^{+54}$ &
5.68 & $10.5_{-1.5}^{+1.8}$ & 134.37/110 \vspace{3pt} \\
\hline
8.4 & 15935 & $754_{-26}^{+29}$ & $>51.49$ & 4.30 &
$6.3_{-1.2}^{+1.2}$ & 102.12/94 \vspace{3pt} \\
& & $1106_{-63}^{+19}$ & $>66.45$ & 2.45 &
$4.8_{-1.5}^{+4.3}$ & \vspace{3pt} \\
\hline
8.5 & 14004 & $728.4_{-8.2}^{+7.9}$ & $55_{-21}^{+37}$ &
4.92 & $7.8_{-1.3}^{+1.7}$ & 55.66/67 \vspace{3pt} \\
& & $1082_{-31}^{+14}$ & $79_{-56}^{+157}$ & 2.41
& $7.4_{-2.1}^{+4.1}$ & \vspace{3pt} \\\hline

	\hline
\end{longtable}
*The upper bound of the width was tied to 78.
\pagebreak
\begin{table*}
	\centering
	\caption{The best-fit spectral parameters for each segment and observation. Errors are quoted at the 90\% confidence level. The superscript `\textit{f}' denotes frozen parameter values.
	}
	\label{tab:table_3}
	\begin{tabular}{lcccccccrrr}
		\hline \hline
		Obs\#. & SXT Exp. & $N_H$ & $\Gamma$ & $kT_e$ & $\tau$ & $T_{in}$ & $T_{bb}$ & \multicolumn{2}{c}{Fe Line @ 6.7 keV} & $\chi^2/dof$ \\
		Seg\# & Time &  &  &  &  &  & & Width & Flux &  \\
		&  & $(10^{22} \, \text{atoms} $  &  & &  &   &  &  & $(10^{-3} \, \text{photons} $ &  \\
		& $(\text{s})$ & $\text{cm}^{-2})$  &  & $(\text{keV})$ &  &  $(\text{keV})$  & (\text{keV}) & (\text{keV}) & $\, \text{cm}^{-2} \, \text{s}^{-1})$ &  \\
		\hline \\ 
	1.1 & 1353 & $2.34_{-0.11}^{+0.12}$ & $1.89_{-0.06}^{+0.06}$ & $3.21_{-0.15}^{+0.18}$ & $11.43_{-0.86}^{+0.87}$ & $0.67_{-0.02}^{+0.02}$ & $<0.42$ & $<$2.0 & $5.3_{-4.4}^{+5.7}$ & 89.05/89 \vspace{3pt} \\
	1.2 & 1355 & $2.36_{-0.12}^{+0.11}$ & $1.89_{-0.06}^{+0.06}$ & $3.22_{-0.14}^{+0.18}$ & $11.35_{-0.84}^{+0.84}$ & $0.65_{-0.02}^{+0.02}$ & $<0.43$ & $1.3_{-1.2}^{+0.7}$ & $6.1_{-4.6}^{+5.5}$ & 113.63/89 \vspace{3pt} \\
	1.3 & 2689 & $2.51_{-0.11}^{+0.08}$ & $1.93_{-0.06}^{+0.06}$ & $3.23_{-0.13}^{+0.14}$ & $10.97_{-0.71}^{+0.82}$ & $0.67_{-0.02}^{+0.02}$ & $<0.48$ & 0.9$^f$ & $<4.5$ & 80.37/94 \vspace{3pt} \\
	1.4 & 1403 & $2.49_{-0.12}^{+0.11}$ & $1.88_{-0.06}^{+0.06}$ & $3.22_{-0.14}^{+0.17}$ & $11.47_{-0.86}^{+0.44}$ & $0.65_{-0.02}^{+0.02}$ & $<0.46$ & $1.3_{-0.6}^{+0.6}$ & $6.8_{-4.2}^{+5.2}$ & 89.62/89 \vspace{3pt} \\
	1.5 & 54.7 & $2.69_{-0.41}^{+0.51}$ & $1.88_{-0.11}^{+0.09}$ & $3.24_{-0.28}^{+0.22}$ & $11.4_{-1.2}^{+1.2}$ & $0.65_{-0.03}^{+0.03}$ & $<0.77$ & $1.1_{-0.9}^{+0.9}$ & $6.4_{-4.8}^{+11}$ & 43.14/65 \vspace{3pt} \\
	1.6 & 1382 & $2.38_{-0.05}^{+0.12}$ & $1.92_{-0.06}^{+0.06}$ & $3.37_{-0.16}^{+0.19}$ & $10.83_{-0.81}^{+0.74}$ & $0.64_{-0.02}^{+0.02}$ & $<0.41$ & $1.4_{-0.8}^{+0.7}$ & $6.7_{-4.4}^{+5.7}$ & 91.60/89 \vspace{3pt} \\
	1.7 & 2085 & $2.42_{-0.10}^{+0.10}$ & $1.94_{-0.06}^{+0.06}$ & $3.43_{-0.16}^{+0.23}$ & $10.54_{-0.83}^{+0.79}$ & $0.64_{-0.02}^{+0.01}$ & $<0.43$ & $1.4_{-1.1}^{+0.9}$ & $4.9_{-4.1}^{+5.6}$ & 103.44/92 \vspace{3pt} \\
	1.8 & 2152 & $2.41_{-0.08}^{+0.14}$ & $1.89_{-0.06}^{+0.06}$ & $3.27_{-0.14}^{+0.17}$ & $11.28_{-0.81}^{+0.80}$ & $0.64_{-0.02}^{+0.01}$ & $<0.43$ & $1.3_{-0.7}^{+0.6}$ & $7.5_{-4.2}^{+5.4}$ & 92.51/93 \vspace{3pt} \\
	1.9 & 1377 & $2.37_{-0.12}^{+0.12}$ & $1.91_{-0.06}^{+0.06}$ & $3.36_{-0.16}^{+0.19}$ & $10.90_{-0.77}^{+0.82}$ & $0.63_{-0.02}^{+0.02}$ & $<0.42$ & $1.3_{-0.6}^{+0.5}$ & $7.4_{-4.1}^{+5.1}$ & 80.95/88 \vspace{3pt} \\
	1.10 & 1382 & $2.34_{-0.08}^{+0.14}$ & $1.95_{-0.06}^{+0.06}$ & $3.48_{-0.16}^{+0.18}$ & $10.35_{-0.51}^{+0.80}$ & $0.63_{-0.02}^{+0.01}$ & $<0.45$ & 0.9$^f$ & $3.6_{-2.7}^{+2.7}$ & 84.68/89 \vspace{3pt} \\
	1.11 & 889.2 & $2.35_{-0.09}^{+0.16}$ & $1.95_{-0.07}^{+0.05}$ & $3.43_{-0.19}^{+0.21}$ & $10.39_{-0.96}^{+0.94}$ & $0.64_{-0.02}^{+0.01}$ & $<0.47$ & $1.4_{-1.2}^{+0.9}$ & $5.5_{-4.4}^{+5.9}$ & 83.33/86 \vspace{3pt} \\
	1.12 & 820.2 & $2.40_{-0.16}^{+0.16}$ & $1.91_{-0.07}^{+0.07}$ & $3.27_{-0.16}^{+0.20}$ & $11.06_{-0.93}^{+0.99}$ & $0.64_{-0.02}^{+0.02}$ & $<0.49$ & $1.1_{-0.7}^{+0.9}$ & $4.7_{-3.5}^{+5.0}$ & 91.89/86 \vspace{3pt} \\
	1.13 & 373.3 & $2.38_{-0.24}^{+0.14}$ & $1.95_{-0.11}^{+0.08}$ & $3.39_{-0.21}^{+0.22}$ & $10.53_{-0.99}^{+1.5}$ & $0.65_{-0.04}^{+0.02}$ & $<0.60$ & 0.9$^f$ & $<6.3$ & 84.63/80 \vspace{3pt} \\
	1.14 & 0\footnote[1]{No SXT exposure time for this segment.} & 2.4$^f$ & $1.87_{-0.09}^{+0.10}$ & $3.23_{-0.17}^{+0.25}$ & $11.5_{-1.3}^{+1.3}$ & $0.62_{-0.03}^{+0.03}$ & $<0.78$ & $1.0_{-0.8}^{+0.9}$ & $6.8_{-4.6}^{+5.5}$ & 1.08/12 \vspace{3pt} \\
	1.15 & 152.2 & $2.09_{-0.12}^{+0.60}$ & $1.90_{-0.05}^{+0.06}$ & $3.39_{-0.17}^{+0.17}$ & $10.90_{-0.21}^{+0.63}$ & $0.62_{-0.03}^{+0.01}$ & $<0.52$ & $1.1_{-0.6}^{+0.7}$ & $5.7_{-4.0}^{+4.9}$ & 68.00/74 \vspace{3pt} \\
	1.16 & 577.6 & $2.36_{-0.11}^{+0.18}$ & $1.94_{-0.07}^{+0.07}$ & $3.35_{-0.17}^{+0.21}$ & $10.68_{-0.95}^{+0.99}$ & $0.65_{-0.02}^{+0.02}$ & $<0.50$ & $1.3_{-1.1}^{+1.1}$ & $4.4_{-3.8}^{+5.1}$ & 79.74/84 \vspace{3pt} \\
	2.1 & 2123 & $2.54_{-0.11}^{+0.10}$ & $1.89_{-0.06}^{+0.06}$ & $3.65_{-0.19}^{+0.25}$ & $10.52_{-0.83}^{+0.86}$ & $0.58_{-0.02}^{+0.02}$ & $<0.44$ & $1.3_{-0.3}^{+0.3}$ & $13.4_{-4.1}^{+4.6}$ & 84.99/89 \vspace{3pt} \\
	2.2 & 2118 & $2.56_{-0.13}^{+0.12}$ & $1.92_{-0.06}^{+0.06}$ & $3.48_{-0.18}^{+0.22}$ & $10.60_{-0.84}^{+0.86}$ & $0.61_{-0.02}^{+0.01}$ & $<0.44$ & $1.4_{-0.4}^{+0.4}$ & $11.8_{-4.4}^{+5.2}$ & 67.66/89 \vspace{3pt} \\
	2.3 & 2347 & $2.50_{-0.11}^{+0.10}$ & $1.89_{-0.05}^{+0.06}$ & $3.35_{-0.15}^{+0.19}$ & $11.09_{-0.80}^{+0.80}$ & $0.62_{-0.02}^{+0.01}$ & $<0.41$ & $1.4_{-0.3}^{+0.3}$ & $12.7_{-4.3}^{+4.9}$ & 93.15/89 \vspace{3pt} \\
	2.4 & 893.9 & $2.53_{-0.15}^{+0.15}$ & $1.95_{-0.07}^{+0.07}$ & $3.61_{-0.21}^{+0.25}$ & $10.15_{-0.86}^{+0.95}$ & $0.61_{-0.02}^{+0.02}$ & $<0.44$ & $1.2_{-0.4}^{+0.4}$ & $10.9_{-4.1}^{+4.7}$ & 83.23/84 \vspace{3pt} \\
	2.5 & 508.8 & $2.70_{-0.17}^{+0.20}$ & $1.97_{-0.08}^{+0.09}$ & $3.66_{-0.21}^{+0.21}$ & $9.8_{-1.2}^{+1.0}$ & $0.61_{-0.01}^{+0.02}$ & $<0.43$ & $1.3_{-0.5}^{+0.4}$ & $9.6_{-4.1}^{+5.1}$ & 83.33/79 \vspace{3pt} \\
	2.6 & 473.1 & $2.37_{-0.18}^{+0.21}$ & $2.02_{-0.08}^{+0.09}$ & $4.23_{-0.33}^{+0.48}$ & $8.7_{-1.0}^{+1.0}$ & $0.60_{-0.02}^{+0.02}$ & $<0.47$ & $1.3_{-0.4}^{+0.4}$ & $10.2_{-4.2}^{+4.6}$ & 90.82/78 \vspace{3pt} \\
	2.7 & 1324 & $2.60_{-0.15}^{+0.13}$ & $1.91_{-0.07}^{+0.07}$ & $3.75_{-0.22}^{+0.29}$ & $10.23_{-0.95}^{+0.97}$ & $0.58_{-0.02}^{+0.02}$ & $<0.46$ & $1.3_{-0.3}^{+0.3}$ & $14.6_{-4.2}^{+4.6}$ & 85.16/86 \vspace{3pt} \\
	3.0 & 2934 & $2.55_{-0.14}^{+0.16}$ & $1.93_{-0.06}^{+0.06}$ & $11_{-3}^{+13}$ & $5.0_{-2.0}^{+1.5}$ & $0.41_{-0.02}^{+0.02}$ & $<0.32$ & $1.1_{-0.1}^{+0.2}$ & $6.3_{-1.3}^{+1.3}$ & 138.23/85 \vspace{3pt} \\
	4.0 & 7591 & $2.48_{-0.11}^{+0.11}$ & $1.92_{-0.04}^{+0.04}$ & $11_{-4}^{+15}$ & $5.1_{-1.8}^{+1.3}$ & $0.39_{-0.01}^{+0.01}$ & $<0.32$ & $1.3_{-0.2}^{+0.2}$ & $6.9_{-1.2}^{+1.2}$ & 140.41/91 \vspace{3pt} \\
	5.0 & 0\footnote[2]{The source was not in the SXT field of view during the observation.} & 2.4$^f$ & $2.00_{-0.07}^{+0.06}$ & $>10.87$ & $4.83_{-0.23}^{+0.28}$ & $0.40_{-0.02}^{+0.02}$ & $<0.59$ & $1.1_{-0.3}^{+0.3}$ & $4.4_{-1.4}^{+1.7}$ & 2.53/14 \vspace{3pt} \\
	6.0 & 2004 & $2.72_{-0.18}^{+0.17}$ & $1.75_{-0.06}^{+0.06}$ & $7.0_{-1.1}^{+2.4}$ & $8.1_{-1.5}^{+1.4}$ & $0.37_{-0.02}^{+0.02}$ & $<0.36$ & $1.1_{-0.2}^{+0.2}$ & $12.8_{-2.1}^{+2.3}$ & 86.96/77 \vspace{3pt} \\
	7.0 & 0\footnote[3]{The SXT PSF image was distorted, showing two bright spots, so the data were not used. } & 2.4$^f$ & $2.16_{-0.09}^{+0.10}$ & $3.87_{-0.29}^{+0.42}$ & $8.3_{-1.0}^{+1.0}$ & $0.63_{-0.01}^{+0.02}$ & $<0.69$ & $1.3_{-0.3}^{+0.4}$ & $8.9_{-3.1}^{+3.1}$ & 1.84/13 \vspace{3pt} \\
	8.1 & 7408 & $2.72_{-0.08}^{+0.08}$ & $1.92_{-0.02}^{+0.04}$ & $3.67_{-0.21}^{+0.20}$ & $10.21_{-0.70}^{+0.39}$ & $0.59_{-0.01}^{+0.01}$ & $<0.36$ & $1.3_{-0.2}^{+0.2}$ & $17.8_{-3.7}^{+3.8}$ & 122.67/95 \vspace{3pt} \\
	8.2 & 8086 & $2.78_{-0.07}^{+0.08}$ & $1.96_{-0.05}^{+0.05}$ & $3.67_{-0.21}^{+0.25}$ & $9.96_{-0.50}^{+0.72}$ & $0.62_{-0.01}^{+0.01}$ & $<0.35$ & $1.4_{-0.2}^{+0.2}$ & $15.9_{-3.9}^{+3.8}$ & 136.83/99 \vspace{3pt} \\
	8.3 & 6674 & $2.68_{-0.08}^{+0.08}$ & $1.95_{-0.04}^{+0.05}$ & $4.32_{-0.29}^{+0.40}$ & $9.07_{-0.40}^{+0.72}$ & $0.56_{-0.01}^{+0.01}$ & $<0.34$ & $1.3_{-0.2}^{+0.2}$ & $16.1_{-3.3}^{+3.3}$ & 144.16/96 \vspace{3pt} \\
	8.4 & 8863 & $2.72_{-0.08}^{+0.08}$ & $1.97_{-0.04}^{+0.05}$ & $3.87_{-0.24}^{+0.31}$ & $9.60_{-0.78}^{+0.66}$ & $0.61_{-0.01}^{+0.01}$ & $<0.36$ & $1.5_{-0.2}^{+0.2}$ & $16.5_{-3.7}^{+3.9}$ & 147.51/98 \vspace{3pt} \\
	8.5 & 3745 & $2.77_{-0.08}^{+0.10}$ & $1.99_{-0.05}^{+0.06}$ & $3.70_{-0.21}^{+0.29}$ & $9.68_{-0.61}^{+0.74}$ & $0.62_{-0.01}^{+0.01}$ & $<0.37$ & $1.3_{-0.2}^{+0.2}$ & $15.3_{-3.7}^{+4.0}$ & 134.89/93 \vspace{3pt} \\

		\hline
	\end{tabular}
\end{table*}
\clearpage
\bibliography{sample7.bib}{}
\bibliographystyle{aasjournalv7}

\end{document}